\documentclass[]{aa}
\usepackage{amsmath}

\usepackage{graphicx}

\usepackage{txfonts}
\newcommand{\bq}{\begin{equation}}
\newcommand{\eq}{\end{equation}}

\begin{document}
   \title{Probing Dark Energy with Baryonic Acoustic Oscillations at
   High Redshifts}

   \author{Ralf S. Koehler
          \inst{1,2}
          \and
          Peter Schuecker\inst{2}
          \and
	  Karl Gebhardt\inst{3}
          }

   \offprints{R. S. Koehler}

   \institute{Universit\"atssternwarte M\"unchen, Ludwig-Maximilians-Universit\"at,
              Scheinerstrasse 1, D-81579 M\"unchen\\
              \email{rkoehler@mpe.mpg.de}
         \and
             Max-Planck-Institut f\"ur extraterrestrische Physik,
             Giessenbachstra{\ss}e 1, D-85748 Garching\\
             \email{peters@mpe.mpg.de}
	\and
	     University of Texas, Astronomy Department, 
	     1 University Station, C1 400, Austin, TX 78712 \\
	     \email{gebhardt@astro.as.texas.edu}
}		
   \date{Received 22.05.2006; accepted 01.09.2006}

% \abstract{}{}{}{}{}
% 5 {} token are mandatory

  \abstract 
% context heading (optional) % {} leave it empty if  necessary 
{To constrain different cosmological models and especially the
equation of state parameter $w$ of dark energy, the phases and amplitudes 
of baryonic acoustic oscillations in
the galaxy power spectrum can be used as cosmological probes.}
% aims heading (mandatory) 
{It will be shown that the phases as well as the amplitudes of 
baryonic acoustic oscillations can be extracted
out of a galaxy power spectrum, including various
observational effects like growth suppression, redshift space
distortions, and galaxy biasing. The phases of baryonic acoustic oscillations
are used as a standard ruler for a cosmological test to
constrain $w$ over a large redshift range with a minimum of assumptions.}
% methods heading (mandatory) 
{A non-oscillating phenomenological fitting function is used to
extract the oscillatory part of the galaxy power spectrum and to disentangle
phase information from amplitude information. The method (FITEX)
is tested with simulated data of the Hubble Volume Simulation to
include redshift space effects, non-linear structure growth and
biasing.  A cosmological test is introduced, which compares the
extracted oscillations against a theoretical model template to derive
constraints on the $w$ parameter.}
% results heading (mandatory) 
{The phenomenological fitting function is found to model the various
distortions of the galaxy power spectrum to the sub-percent level. The various distortions
only boost the amplitude of the oscillations. The
theoretical template is accurate enough to test for small deviations
in the phases of the oscillations, resulting from different $w$ values. A
cosmological test, using the baryonic acoustic oscillations
as a standard ruler, is able to constrain $w$ in a robust way. These constraints are based
solely on the phase of these oscillations, assuming no errors of the value
at the end of the Compton drag epoch, as the standard ruler. The $w$ constraints 
are expected to further improve when the full two dimensional redshift space effects
and the amplitudes of the baryonic acoustic oscillations are taken into account.}
% conclusions heading (optional), leave it empty if necessary {}


   \titlerunning {Probing Dark Energy with Baryonic Acoustic Oscillations}

   \maketitle
%
%________________________________________________________________

\section{Introduction}\label{INTRO}

Acoustic oscillations as observed in the temperature anisotropies of
the cosmic microwave background radiation (CMB) are traditionally used
to constrain the values of certain cosmological parameters. The
intrinsic amplitudes and locations of the oscillations are determined
by the densities and pressures of the various energy components in the
very early and hot Universe. The oscillations are furthermore modified
by their subsequent geometric projection onto the present hypersphere,
where they are observed. It turns out that both intrinsic CMB
oscillations and their geometric projection are mainly shaped by the
physical properties at high redshifts. Excluding the very large scales
where cosmic variance makes analysis quite difficult, the observed
oscillations in the CMB are highly degenerate against certain changes of the
energy density $\rho_{\rm DE}$ and the equation of state parameter,
$w=p_{\rm DE}/\rho_{\rm DE}c^2$, of the dark energy, dominating at low
redshifts (see also Caldwell, Dave \& Steinhardt 1998).

However, related so-called baryonic acoustic oscillations (BAOs) are
expected to be observable at lower redshifts in the matter power
spectrum. Therefore, the phases of the BAOs can be used as cosmic
rulers in a similar manner as the CMB oscillations, but now at much
smaller redshifts (Eisenstein, Hu \& Tegmark 1998). Here, the observed
phases are affected by the late-time geometry and thus by the value of
$w$, the prime cosmological parameter of the present investigation.

The BAOs are classical Doppler peaks in the density
distribution of matter (e.g. Hu \& Sugiyama 1996). They are triggered in
the hot Universe by oscillatory velocity patterns of the
baryons. These velocity oscillations are generated, in the same
environment as the CMB density oscillations, by sound waves on scales
where radiation pressure could stabilize the fluid against
gravitational collapse. At later times, this oscillatory velocity
pattern of the baryon field produces kinematically a new field of
matter density fluctuations in form of quasi-regular matter
oscillations. The amplitudes of these BAOs are determined by the
ratio of baryonic matter to the overall matter density.  The BAOs thus constitute a quasi-regular
pattern of oscillating substructures superposed with small amplitudes
onto a general density field which fluctuates irregularly with
several 100 times higher amplitudes.

The imprints of the BAOs on galaxy and clusters distributions,
gravitational shear maps etc. can provide at least in principle a
clean cosmic ruler for precise tests of the $w$ parameter. The
characteristic scale, $s$, of these imprints is the comoving distance,
sound waves can travel during the epoch where baryons and photons are
strongly coupled through Compton and Thomson scattering (Compton
drag). Cosmological tests based on the resulting sound horizon, $s$,
as a metric ruler are expected to have only very small systematic
errors because the phases of the BAOs located at scales small compared
to $s$ are solely determined by well-understood physical processes. A
practical problem is, however, to separate the oscillatory BAO modes
from the irregular fluctuation field with sufficient accuracy.

One might object that the small amplitudes of the BAOs can easily be
washed out especially by structure growth which can mix perturbation
modes with adjacent wavenumbers $k$. This is certainly true for
oscillations in the non-linear regime (Meiksin, White \& Peacock
1999). On larger scales, however, recent observations of 2dF galaxies
at redshifts $z<0.3$ clearly show several BAOs in the galaxy power
spectrum (Cole et al. 2005). In addition, the sequence of BAOs in
$k$-space is projected into a single wiggle in the space domain. The
corresponding excess galaxy correlation was in fact observed in the
two-point spatial correlation function of the luminous red SDSS
galaxies at $z<0.5$ (Eisenstein et al. 2005). In addition to these
observational indications, results from recent numerical simulations
suggest that within certain redshift and scale ranges, stable BAOs
could exist as useful probes for cosmological investigations (e.g. Seo
\& Eisenstein 2005, Springel et al. 2005, Jeong \& Komatsu 2006). However, several 
critical issues like the exact behavior of baryons during structure
growth on BAO-scales can only be discussed with much larger
simulations and a better understanding of galaxy formation.

Applications of BAOs for cosmological tests of $w$ are confronted with
the following situation. Present observational constraints of $w$
using different combinations of CMB, galaxy clusters, galaxies, and
gravitational lensing data are all found to be consistent with $w=-1$,
i.e., the value for the cosmological constant. The $1\sigma$ error
of the $w$ values, derived from partially dependent observations,
is around 10-20\%
(for a recent review see, e.g., Schuecker 2005). To improve current
estimates, we thus have to extract the BAOs and to measure their
phases quite accurately in the presence of non-linear and
scale-dependent effects. Therefore, even tiny systematic errors in the
analysis on percent levels can have severe consequences for the
observational accuracy of $w$.

The basic aim of the present paper is to describe a new, simple, and
robust method to extract BAOs from (non)linear, scale and
redshift-dependent biased, redshift-space galaxy power spectra under
realistic lightcone survey conditions. For brevity, we call the method
`fit and extract' (FITEX). The method has the potential to reduce many
of the abovementioned sources of systematic errors. Our starting point
is to use as few assumptions as possible for the extraction of the
BAOs from a complex power spectrum. For reasons which will become
clear later, we do the separation by fitting a flexible
non-oscillating function to an observable which is directly related to
the transfer function and not to the power spectrum itself. The
crucial point is to show that BAOs extracted in this simple manner are
still governed by simple physical processes, and that FITEX stays
robust even under complex survey conditions. We believe that such
model-independent approaches are quite important, especially in light
of the fact that at least in the near future we cannot expect to model
the abovementioned non-linear and scale-dependent effects with
accuracies on the sub-percent level.

Our computations are in several cases optimized to galaxy surveys
covering typically several 100 square degrees on the sky in the
redshift range between $z=2-4$. This choice is motivated by the Hobby
Eberly Dark Energy Experiment (HETDEX), which is planned to measure
the $w$ parameter with several million Ly-$\alpha$ emitting galaxies
at these redshifts (Hill \& McQueen. 2004, Gebhardt et al. in prep.). We see
the results of the present paper as a useful contribution to a
realistic error forecast for this important project.

We organized the paper as follows. In Sect.\,\ref{BAOS} we motivate
the basic idea of the FITEX algorithm. In the following sections we
test its performance to separate BAOs from a complex power spectrum
under linear conditions (Sect.\,\ref{EXTRACT}), under quasi-non-linear
conditions (Sect.\,\ref{NONLINEAR}), in redshift space
(Sect.\,\ref{REDSHIFT}), and from biased samples
(Sect.\,\ref{BIASING}). Finally, we illustrate the performance of
FITEX including all effects and use a cosmological test of $w$
as a benchmark to investigate the quality of the BAO extraction together with the
theoretical template (Sect.\,\ref{CTEST}). Our results are mainly based on a deep data
wedge extracted by the Virgo Consortium from the Hubble Volume
Simulation. The paper thus presents for the first time constraints on
the application of BAOs for cosmological tests under realistic
lightcone survey conditions. 

If not mentioned explicitly, we assume a spatially flat
Friedmann-Lema\^{i}tre Robertson-Walker world model with the Hubble
constant in units of $h=H_0/(100\,{\rm km}\,{\rm s}^{-1}\,{\rm Mpc^{-1}})$,
the present values of the total matter density $\Omega_{\rm
m}h^2=0.147$ and baryon density $\Omega_{\rm b}h^2=0.0196$, the density of
relativistic matter (e.g. neutrinos) $\Omega_\nu =0$, and the
mean CMB temperature $T_{\rm CMB}=2.728\,{\rm K}$ (concordance cosmology).

%__________________________________________________________________

\section{Baryonic Acoustic Oscillations (BAOs)}\label{BAOS}

In this section, we discuss the `theoretical wiggle function', that
is, a reference function (see Eq.\,\ref{eq wiggles} below) we use in our
cosmological tests of $w$ to match the BAOs extracted in a certain
manner from the complex galaxy power spectrum. The power spectrum may
be written in the form $P(k,z)\,=\,A(k,z,b)\,k^n\,T^2(k)$. The
amplitude $A(k,z,b)$ includes the primordial amplitude, redshift and scale-dependent effects of
redshift space distortions, linear and non-linear structure growth,
and galaxy biasing which will be discussed in the course of the
paper. The exponent $n$ is the slope of the primordial power spectrum,
and $T(k)$ the transfer function. Our further treatment of the
formation of BAOs is based on Hu \& Sugiyama (1996) where a detailed
description of the relevant physical processes can be found, and on
the fitting equations derived by Eisenstein \& Hu (1998).

Between the epoch of matter-radiation equality at
redshift $z_{\rm eq}\approx3526$ (concordance cosmology) and the end
of the Compton drag epoch at $z_{\rm d}\approx1026$, the baryons follow a regular velocity
pattern which can generate new density fluctuations kinematically.  On
small scales and at $z_{\rm d}$, this effect (velocity overshoot)
overrides the intrinsic density fluctuations of the baryons. As
adiabatic modes dominate the isocurvature modes, the baryon density
oscillates as $\sin(ks)$. The phase of the BAOs is frozen out at
$z_{\rm d}$ at the value $ks$, with $s\approx152\,{\rm Mpc}$ being the
comoving sound horizon at $z_{\rm d}$ and $k$ the comoving wavenumber
of the oscillation. The BAOs are thus $\pi/2$ out of phase with the
corresponding CMB fluctuations. The sound horizon $s$ at $z_{\rm d}$
is the standard ruler we are looking for. Its value 
calibrates the phases of the theoretical wiggle function and can be 
measured quite easily with CMB experiments on the sub-percent 
level. This includes the contribution of unknown relativistic 
energy components (Eisenstein \& White 2004).
For consistency with the Hubble Volume Simulations 
(see Sect.\;\ref{NONLINEAR}) we use $\Omega_{\rm \nu} = 0$.

The sinusoidal fluctuations in the baryon density are dampened by
expansion drag, gravitational forcing, and Silk damping. In addition,
cosmic expansion forces the velocity contributions to fall off at
large scales, and amplitudes decline when dark matter dominates the
energy density. While these effects only reduce the amplitude of the
fluctuations and do not affect the phase, the damping may be summarized by the
pseudo transfer function $T_w(k)\sim j_0(k\,s)e^{-(k/k_{\rm
Silk})^{m_s}}/[1\,+\,(\beta_b\,/\,k\,s)^{3}]$, with $j_0$ the
spherical Bessel function of order zero, the exponent $m_{\rm
s}\,\approx\, 1.4$ which is basically independent from cosmology, $
k_{\rm Silk}=1.6(\Omega_{\rm b}h^2)^{0.52}(\Omega_{\rm m}h^2)^{0.73}
[1+(10.4\Omega_{\rm m}h^2)^{-0.95}]$, and $\beta_b=0.5+f_{\rm
b}+(3-2f_{\rm b})\sqrt{(17.2\Omega_{\rm m}h^2)^2+1}$ with $f_{\rm
b}=\Omega_{\rm b}/\Omega_{\rm m}$. However, velocity overshoot
dominates only on scales small compared to $s$. On larger scales, the
original sound horizon, $s$, appears to be reduced by
$(\tilde{s}/s)^3=1+(\beta/ks)^3$, with $\beta=8.41(\Omega_{\rm
m}h^2)^{0.435}$. This correction of $s$ is about 0.2\% at
$60\,h^{-1}\,{\rm Mpc}$ and 1.3\% at $120\,h\,^{-1}\,{\rm
Mpc}$. Though relatively small, the corrections are important as they
change the length of the standard ruler $s$ depending on
$k$. Collecting all scale-dependent terms we get the oscillatory
solution (theoretical wiggle function)
\bq
    \label{eq wiggles}
    T_w(k)\;\sim\; \frac{\,e^{-(k/k_{\rm
    Silk})^{m_{\rm s}}}}{1\,+\,(\beta_b\,/\,k\,s)^3}\,j_0(k\,\tilde{s})\,.
\eq

A way to extract BAOs from a complex power spectrum can be found when
we specify the relation between the wiggle function, $T_w(k)$, and the
total transfer function $T(k)$, which we discuss now. Each particle
species, in the present case CDM and baryonic matter with the
corresponding densities $\Omega_{\rm c}$ and $\Omega_{\rm b}$, should
have separate effective transfer functions, $T_c$ and $T_b$. Though,
after the drag epoch at $z_{\rm d}$, baryons appear basically
pressureless and will fall into the potential wells of CDM. This
results in a transfer function valid for both species of matter,
\bq\label{TK}
T(k)\,=\,\frac{\Omega_{\rm c}}{\Omega_{\rm m}}T_c(k)\,+\,
\frac{\Omega_{\rm b}}{\Omega_{\rm m}}\,T_b(k)\,,
\eq
with
$T_c(k)=f\tilde{T}_0(k,1,\beta_c)+(1-f)\tilde{T}_0(k,\alpha_c,\beta_c)$,
written in terms of the generalized transfer function
$\tilde{T}_0(k,\alpha_c,\beta_c)$ as defined in Eisenstein \& Hu
(1998). Here, $f=1/[1+(ks/5.4)^4]$ smooths the combination of the
almost baryon-free and baryon-loaded solutions near $s$, and
\bq\label{T_b}
T_b(k)\,=\,\tilde{T}_{\rm b}(k)\,+\,T_w(k)\,,
\eq
where $\tilde{T}_{\rm
b}(k)=\frac{j_0(k\tilde{s})\tilde{T}_0(k,1,1)}{1+(ks/5.2)^2}$.  These
approximations are better than 2\% for $\Omega_{\rm b}\,/\,\Omega_{\rm
m}\,<\,0.5$. Note that $\Omega_{\rm b}/\Omega_{\rm m}\rightarrow 0$
corresponds to $\alpha_c,\beta_c\rightarrow 1$.  If $\Omega_{\rm
c}\,>>\, \Omega_{\rm b}$, the effects of dark matter dominate over
velocity overshoot of the BAOs.  Written in this way, we immediately
see that the oscillatory part (\ref{eq wiggles}) is added on top
of a non-oscillatory part and can be computed by subtracting off a
smooth continuum from the complete transfer function. This process
establishes the basic methodology of FITEX.

\begin{figure}[htb]
    \centering \includegraphics[width=8.8cm]{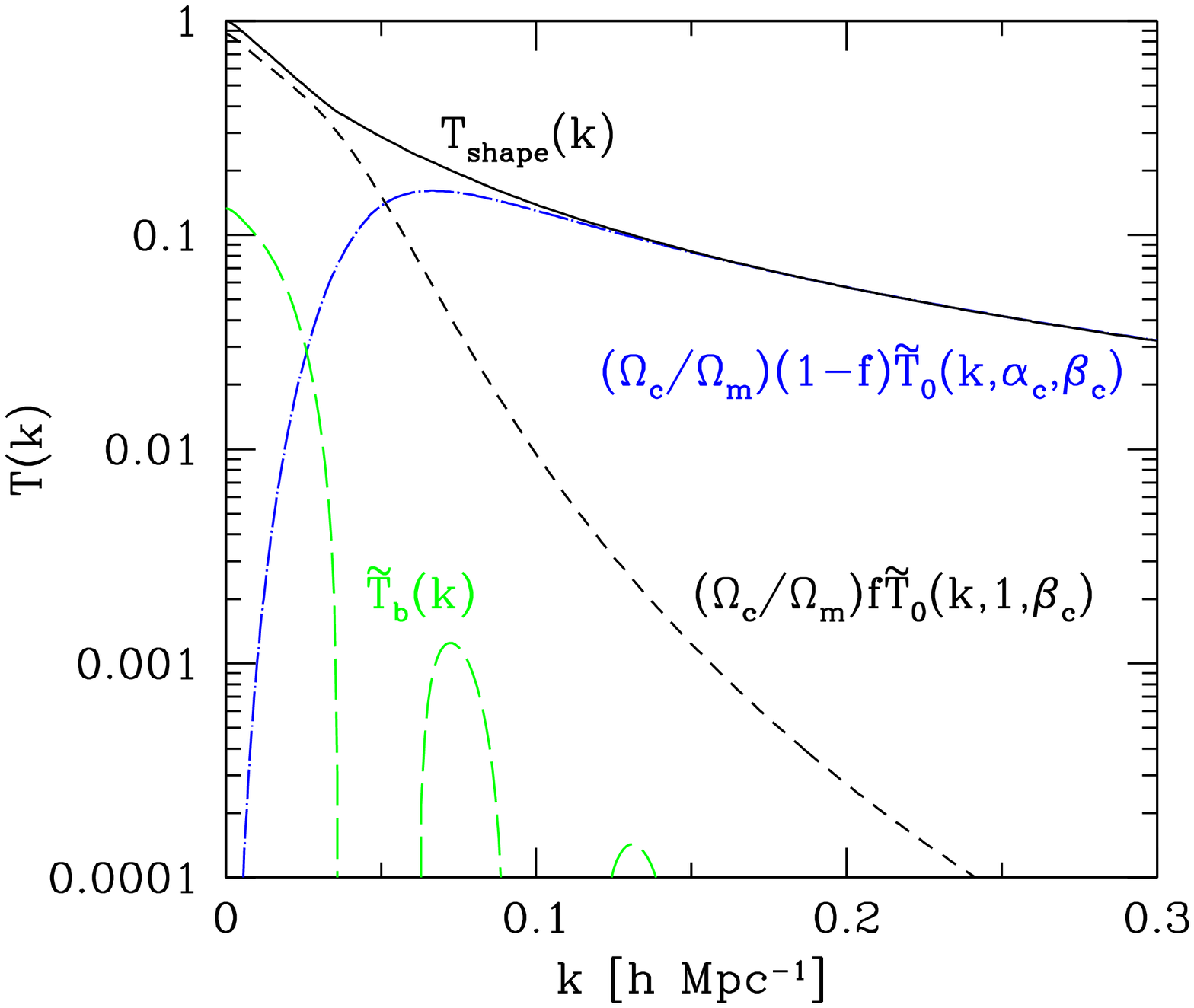}
    \centering \includegraphics[width=8.8cm]{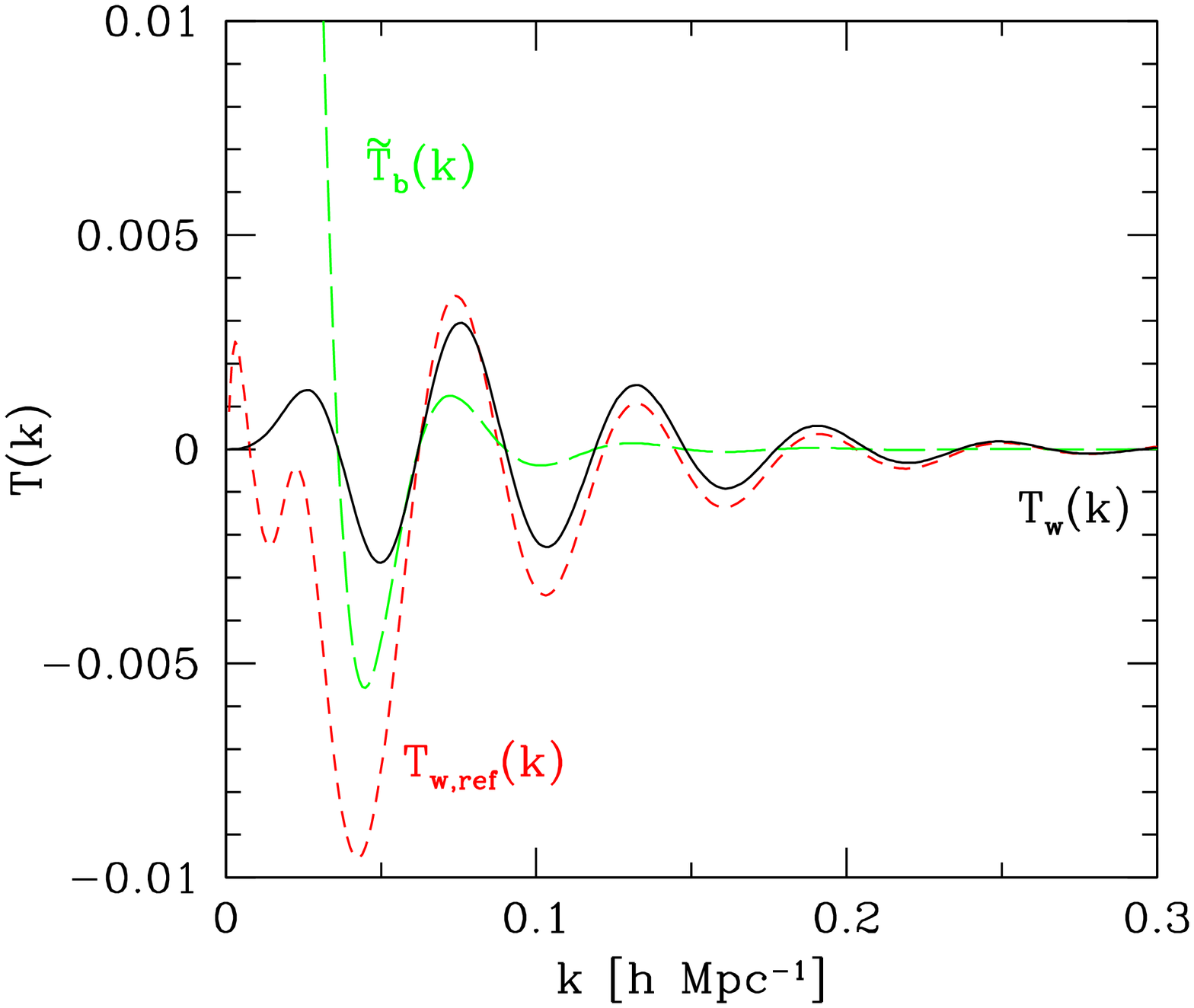}
    \caption[Transfer function] {\small Comparison of different terms
    of the transfer function computed with $\Omega_{\rm
    b}\,h^2\,=\,0.0196$, $\Omega_{\rm m}\,h^2\,=\,0.147$ and $T_{\rm
    CMB}\,=\,2.728$\,K. {\bf Upper panel:} Non-oscillatory parts. The
    function $T_{\rm shape}$ is the sum of the other three. {\bf Lower
    panel:} Oscillating parts. The residuals of the subtraction 
    between the transfer function and its corresponding non-oscillatory 
    fit is shown as long-dashed line. Note the different scaling of the 
    y-axes. See main text for more details.} \label{fig transfer}
\end{figure}

The different terms of the transfer function are shown in
Fig.\,\ref{fig transfer}. The upper panel shows the components which
determine the global shape of the transfer function. Note that an
additional oscillating component, the first term in Eq.\,(\ref{T_b}),
is necessary to describe the transition from large to small scales
with sufficient accuracy.  This is a direct consequence of the fact
that the main effect of the baryons on the transfer function is a
damping of the overall growth of dark matter between $z_{\rm eq}$ and
$z_{\rm d}$ which significantly reduces the fluctuation power at
scales smaller than $s$. The resulting break is the imprint of 
the baryons on the power spectrum which is easiest to observe. The term
$\tilde{T_b}$ thus contributes mainly to the shape as it is rapidly
declining on scales smaller than $s$ to model the damping effect of the baryons.

In the lower panel of Fig. \ref{fig transfer}, the two parts of the
baryon transfer function are shown. One can clearly see the first to
fifth oscillations. Note that $\tilde{T_b}$ declines very fast and is
almost zero after the second oscillation. It is compared to the second
term of the baryonic transfer function, $T_w(k)$, and a reference
function, $T_{w,\textrm{ref}}$, which is formally constructed to contain
all oscillations. Thus, the reference function includes both
the small-scale oscillations from $T_w(k)$ and the large-scale
oscillations from $\tilde{T}_{\rm b}(k)$. The latter is, however, also
determined by the dark matter function $T_0(k,1,1)$ (see below
Eq.\,\ref{T_b}). Therefore, excluding or including knowledge of the
shape of the dark matter function for the extraction of BAOs
corresponds to working either with $T_w(k)$ or with
$T_{w,\textrm{ref}}$. Both $T_w(k)$ and $T_{w,\textrm{ref}}$ agree
reasonably well beyond $k\,\geq\,0.05\,h\,\textrm{Mpc}^{-1}$,
especially when one keeps in mind that only the phases of the
oscillations are relevant as the amplitudes are subject to distortions
anyway (see below).

The good agreement between the phases of $T_w(k)$ and $T_{w,\textrm{ref}}$ means,
that quite simple physics as described by $T_w$ (mainly velocity
overshoot and Silk damping) dominate the oscillations expected to be
seen in the overall power spectrum. However, on larger scales, the physics get
more complex due to the growing influence of CDM and the backreaction
of the baryons on structure growth. Thus $T_w$ and
$T_{w,\textrm{ref}}$ differ accordingly. Nevertheless, on small
scales, $T_w$ is a good theoretical wiggle function for BAOs phases which can
be used as a standard ruler for cosmological tests, with a minimum of
theoretical assumptions.

\section{Extracting BAOs under Linear Conditions}\label{EXTRACT}

The main advantages of constraining cosmological parameters with BAOs are its small
systematic errors, the ability to discriminate between geometrical effects (homogeneous universe) and structure growth effects (inhomogeneous universe) of cosmological parameters, and the potential to
constrain dark energy without assuming a certain dark matter model.
The observed phase of the BAOs is only affected by the geometry of the universe, while the observed amplitude of the oscillations is mainly affected by structure growth and other amplitude effects (see Sects. \ref{NONLINEAR} to \ref{BIASING}). It is thus very important to extract the BAOs from the power spectrum in a way that does not mix these two benchmarks.
As suggested by the previous discussion, BAOs, described by simple 
physical processes, should be extracted by \emph{subtracting} the shape of the
non-oscillatory part of the \emph{transfer} function. This method is able to disentangle phase effects from amplitude effects in an effective manner.
 
In contrast to this method, previous studies (Blake \& Glazebrook 2003, Angulo et al. 2005, H\"utsi 2005, Seo \& Eisenstein 2005, White 2005) used the CDM-based theoretical non-oscillatory 
\emph{power spectrum}, $P_{\rm ref}(k,z)$, given in Eisenstein \& Hu (1998) to analyze the BAOs. 
This is done by \emph{dividing} the measured power spectrum, $P_{\rm obs}(k,z)$, by this 
reference power spectrum. Assuming that the transfer function
can be split into a shape part, $T_{\rm shape}$, and an oscillatory
part, $T_{\rm wiggle}$, where $T_{\rm shape}\,>\,T_{\rm wiggle}$
(Sect.\,\ref{BAOS}), the result of such a division is
\bq
    \label{eq_division}
    \begin{split} \frac{P_{\rm obs}(k,z)}{P_{\rm
    ref}(k,z)}\;&=\;\frac{A_{\rm obs}(k,z)\,[T_{\rm
    shape}(k)\,+\,T_{\rm wiggle}(k)]^2\,k^n} {A_{\rm ref}(k,z)\,T_{\rm
    shape}^2(k)\,k^n}\;\\ &\approx\;2\, \frac{A_{\rm obs}(k,z)\,T_{\rm
    wiggle}(k)}{A_{\rm ref}(k,z)\,T_{\rm shape}(k)}\, +\,\frac{A_{\rm
    obs}(k,z)}{A_{\rm ref}(k,z)}\;. \end{split}
\eq
Eq.(\ref{eq_division}) gives a spectral ratio, which strongly deviates from the simple 
functional form of Eq.\,(\ref{eq wiggles}) expected from basic physical principles, 
and thus unnecessarily complicates any further analysis of the resulting ``BAOs''.
A possible result of such a division is given in Springel et al. (2005, Fig. 6). 
It can easily be seen that phase information and amplitude information is mixed.

In addition, the use of a \emph{theoretical} non-oscillatoy reference power spectrum, $P_{\rm ref}(k,z)$, 
or the related \emph{theoretical} ``boosted'' transfer function, 
$A_{\rm ref}(k,z) T_{\rm ref}(k,z) = \sqrt{P_{\rm ref}(k,z)/k^n}$, 
has a number of drawbacks:

(1) The shape of the transfer function and thus the power spectrum vary with the cosmological
parameters. To compute a reference power spectrum one has to
know the exact values of several cosmological parameters.  Furthermore
one has to assume a CDM model as a prior. 
(2) The reference power spectrum
has to be flexible to compensate for various distortions: structure
growth, redshift space distortions, biasing etc. distort the transfer
function and thus its shape. Most of the effects vary with redshift as
well as with $k$. 
(3) The reference power spectrum has to be exact on the
sub-percent level. The BAOs make up only $\sim 2\%$ of the transfer
function depending on the ratio $\Omega_{\rm b}/\Omega_{\rm m}$. An
analytical function has to be more accurate to extract the
oscillations from the transfer function. 
(4) It is non-trivial to provide an assumption-free template for the 
resulting power ratio to perform a cosmological test. Instead, very accurate
knowledge of the cosmology as well as amplitude effects is needed.

Most of these issues are well-known and useful corrections are already
available from the simulations. However, these calibrations have
certainly not reached the sub-percent accuracy level which is
necessary for the cosmological tests, and one can doubt whether this
is achievable at all. Furthermore, one of the main advantages of the BAO method,
the ability not to assume a certain dark matter model, is negated.

The Eqs.\,(\ref{eq wiggles}-\ref{T_b}) instead suggest that
the computation of the difference of transfer functions to extract BAOs is the
most direct approach. Therefore, FITEX does not \emph{divide} the observed \emph{power spectrum} by
a \emph{theoretical} non-oscillating reference power spectrum but \emph{subtracts} a \emph{phenomenological} 
non-oscillating fitting function from the observed boosted \emph{transfer function}, to extract
the BAOs. Formally FITDEX can be described as follows:

\begin{equation}\label{FITEX}
\sqrt{\frac{P(k,z,b,)}{k}}-F(k,b,z)=\sqrt{A(k,z,b)}
\frac{\Omega_{\rm b}}{\Omega_{\rm m}}T_{\rm w}(k)\,.
\end{equation}
We found that the formula
\bq
    \label{eq fit}
    F(k)\;=\; \frac{A}{1\,+\,B\,k^\delta}\,e^{(k\,/\,k_1)^\alpha}\;
\eq
is able to fit all non-oscillating distortions to a satisfactory 
degree and leaves enough free parameters to allow for a wide range of 
transfer functions.
Eq.\,(\ref{eq fit}) has no oscillatory components and can thus only
trace the shape of a transfer function.
Most of the drawbacks of the reference power spectrum division can be avoided:

(1) No cosmology or dark matter model has to be assumed to extract the BAOs.
(2) The formula is flexible enough to compensate for various non-oscillating distortions without having 
to care about the source or physics of the distortions (see Sect. \ref{NONLINEAR} - \ref{BIASING}).
(3) The formula is able to fit power spectra reliably on the sub percent level (see Sect. \ref{NONLINEAR}).
(4) It is possible to provide a theoretical template function that can be used without assuming
knowledge about dark matter and various amplitude effects by leaving the amplitude as a free parameter, as phase information and amplitude information is disentangled. The amplitude could be used for a further cosmological test.

\begin{figure}[htbp]
    \centering \includegraphics[width=9cm]{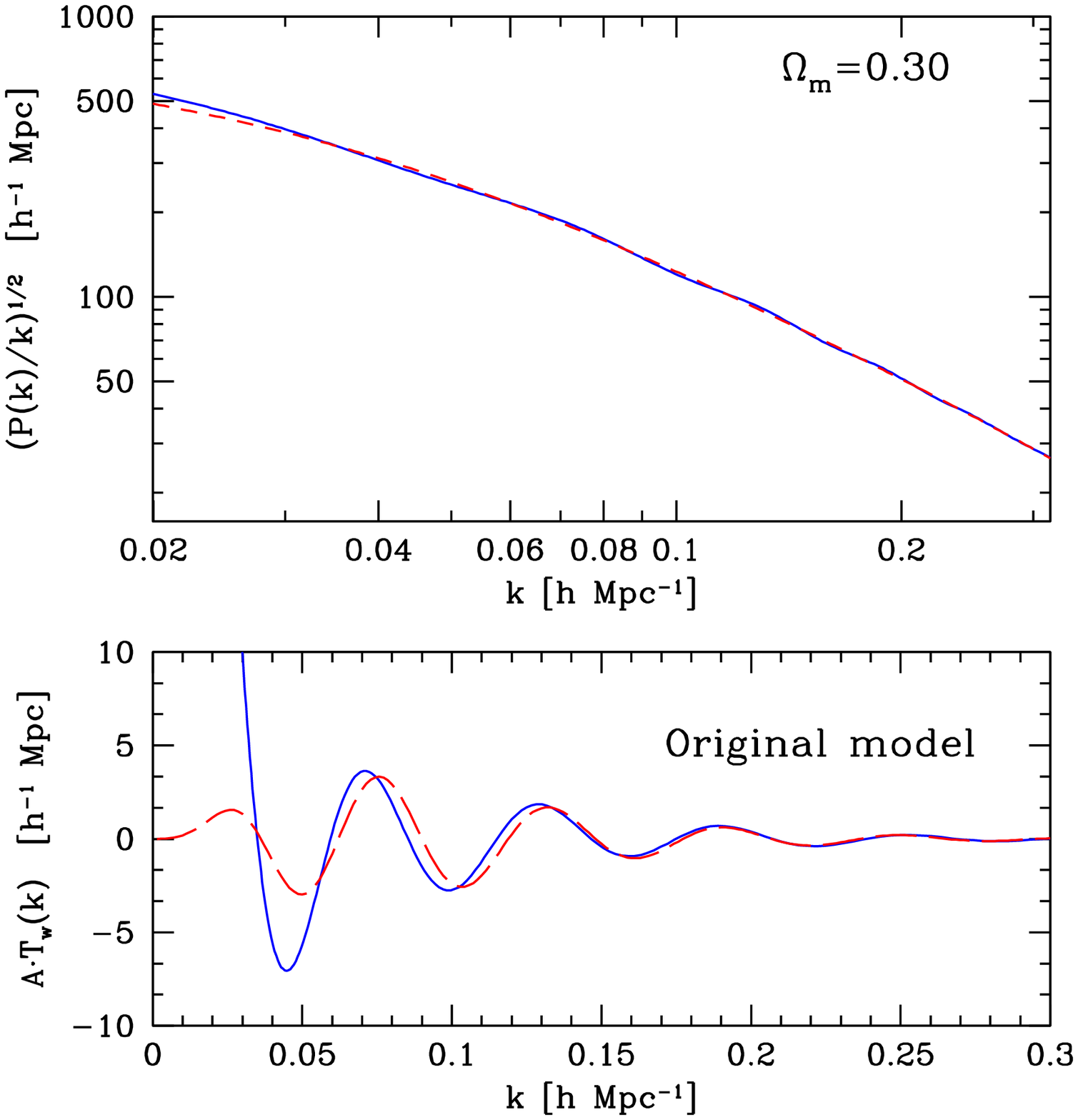}
    \centering \includegraphics[width=9cm]{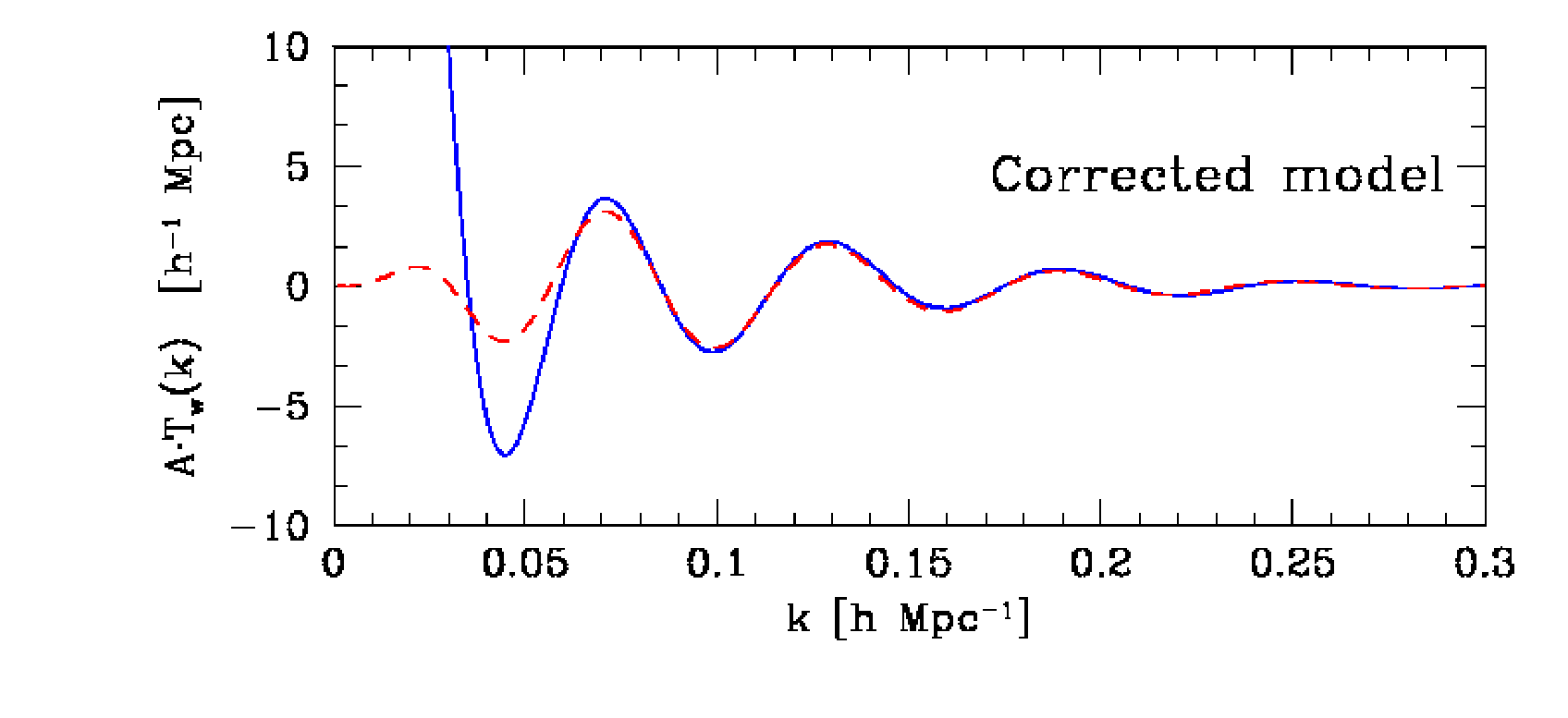}
    \caption[Template corrections] {\small Performance of the
    phenomenological fitting function under linear conditions. The upper panel shows the CMB fast transfer function (solid line) and the best non-oscillating fit (dashed line). In the lower panels, the residuals of the subtraction are plotted (solid line) as well as the theoretical transfer function (dashed line).
The original model gives the performance of the original theoretical function (Eq. \ref{eq wiggles}) while the corrected model uses the modified function (Eq. \ref{eq correction})}
    \label{fig tempcorrect}
\end{figure}

To test FITEX, i.e., the combination of both the fitting function
$F(k)$ and the wiggle function $T_w(k)$ under linear conditions, we
first computed normalized transfer functions with CMBfast with the parameter values $w=-1$ (the transfer functions are affected by $w$ only on scales above several Gpc, Ma et al. 1999), $\Omega_{\rm m}=0.3$, $\Omega_{\rm b}=0.04$, $h=0.70$, the primordial slope $n=1$, the average CMB temperature $T_{\rm CMB}=2.728$\,K, the He mass fraction after primordial nucleosynthesis $Y_{\rm He}=0.24$, and the number of neutrino families $N_\nu=3.04$. In the
second step, the phenomenological continuum function (Eq.\,\ref{eq
fit}) was fitted to the multi-component transfer functions in the
range $0.01<k<0.3h\,\,\textrm{Mpc}^{-1}$. The results were subtracted
from the original transfer functions to extract the BAOs. One example
with $\Omega_{\rm m}=0.3$ is shown in Fig.\,\ref{fig tempcorrect}. In
the first  panel, the multi-component transfer function is
plotted (solid line) as well as the phenomenological fitting
function (dashed line). In the second and third panels, the
differences between the transfer function and the continuum fit are
plotted (solid lines), together with the theoretical wiggle functions (dashed lines)
computed with Eq.\,(\ref{eq wiggles}).

With typical baryon fractions, $f_b\sim0.15$, baryonic oscillations
are hardly visible as they make up only $\sim2\%$ of the
multi-component transfer function. This illustrates the main
observational challenge in using wiggles as a metric ruler for
cosmological investigations. On small scales,
$k>0.15h\textrm{Mpc}^{-1}$, Fig. \ref{fig tempcorrect} shows that the
theoretical wiggle function describes the baryonic oscillations to a
satisfactory degree. This is expected, as the transfer function is
dominated by the effects of velocity overshoot and Silk damping on
these scales and the wiggle function was designed to model these
effects. On large scales, $k<0.05\,h\textrm{Mpc}^{-1}$, the BAOs have
to cope with competing CDM effects that dominate in these
$k$-ranges. As the influence of CDM increases, the transfer function
can not be described by baryonic physics alone and further corrections
and assumptions have to be applied. Two related effects are:

(1) Turnover-effect: The turnover in the matter power spectrum occurs
on the scale $k_{\rm eq}$ of the particle horizon at $z_{\rm eq}$,
which coincides with the approximate location of the first wiggle. No
observation has hitherto clearly revealed this turnover in spectral
power. The main problem in identifying the turnover is mostly due to
the fact, that it is located at large scales where precise
measurements of spectral power are difficult: the surveys cover too
small volumes with slice-like shapes leading to strong smoothing (up
to factors of two for the 2dF survey, Percival et al. 2001) and significant leakage in the
derived power spectra. Under such conditions it appears quite
difficult if not impossible to discriminate the first wiggle from the
turnover when even the position of the turnover could not yet been
determined to some accuracy. To do this, an excellent signal to noise
ratio would be needed as well as precise knowledge of the underlying
linear theory matter power spectrum. This includes information of
distortions of the power spectrum to the sub-percent level. As a
consequence, a model-independent approach like separating the BAOs
from the multi-component transfer function with a phenomenological
fitting function will only be able to detect the first wiggle when the
baryon fraction is about as high as 40 percent or more, as we found
with CMBfast simulations. However, we will show that our cosmological
test is most sensitive to changes of $w$ on scales around
$k=0.1h\textrm{Mpc}^{-1}$. Therefore, the inability of a
model-independent approach to detect the first BAO does not matter
much as this first wiggle barely contributes any information to the
cosmological test, even more as the amplitude of the first wiggle is
comparatively low as it is dampened by CDM effects (see Sect. \ref{CTEST}).

(2) Phase-shift effects: As mentioned in Sect.\,\ref{BAOS}, the
increasing effect of dark matter on large scales introduces a
$k$-dependent phase-shift of the BAOs. The theoretical wiggle function
(Eq.\,\ref{eq wiggles}) already includes a phenomenological correction
for this effect (replacing $s$ by $\tilde{s}$). A similar effect on
the phases, but in the opposite direction, is introduced by the
damping of baryons on structure growth. In fact, the main effect of
baryons on the multi-component power spectrum is, assuming a standard
cosmology, a sharp break in power starting on scales smaller than the
turnover. Formally, this break is caused by the rapid decay of the
$\tilde{T}_b$ component of the transfer function (see Fig. \ref{fig
transfer}). As the break occurs right after the turnover, it is very
hard to distinguish shape components from oscillations. For the
extraction of BAOs in that scale range we have found that this
spectral break must be modeled in detail (with an accuracy on the
sub-percent level) or artificial phase-shifts with sizes of the order
of interesting $w$ effects are introduced when naive fitting functions
ignoring this baryonic backreaction are subtracted from the
multi-component power spectrum. Thus the model independent fitting
function presented in this work (Eq.\,\ref{eq fit}), but also
traditional wiggle-free dark matter transfer functions with a slightly
different functional approach like those given in Bardeen et
al. (1986) or Efstathiou, Bond \& White (1992), introduce artificial
phase-shifts of the generic form
\bq
    \label{eq correction}
    \left(\frac{\hat{s}}{\tilde{s}}\right)^{3/2}=\;\,0.98\,+\,
    \left[\frac{5.1\,(\Omega_{\rm m}\,h^2)^{0.47}}
     {k\,s}\right]^{3/2}\;,
\eq
which gives a useful summary of the phase-shifts (about 0.7\% at
$60h^{-1}{\rm Mpc}$ and 4\% at $120h^{-1}{\rm Mpc}$) even for very
different baryon fractions. This equation was derived phenomenologically 
from fits to CMBfast simulations. The second panel in Fig.\,\ref{fig
tempcorrect} shows an example for the concordance cosmology. It is
seen that the phase-shifts are most prominent at intermediate scales,
$k\,<\,0.15\,h\,\textrm{Mpc}^{-1}$, and counteracts the phase-shifts
introduced by large-scale CDM potentials. The dashed line in the
fourth panel shows the corrected theoretical wiggle function,
replacing $\tilde{s}$ in Eq.\,(\ref{eq wiggles}) by $\hat{s}$ from
Eq.\,(\ref{eq correction}). It is clearly seen that the extracted
wiggle function follows the predicted theoretical, and
phenomenologically corrected, wiggle function on the 0.5\%-level in
the range $0.05<k<0.3\,h\,\textrm{Mpc}^{-1}$ which is relevant for a
cosmological test of the parameter $w$. In the following,
Eqs.\,(\ref{eq wiggles}) and (\ref{eq correction}) constitute our
final theoretical wiggle function which we use in FITEX to match the BAOs
extracted from the simulated data.

\section{Extracting BAOs under Quasi-Nonlinear Conditions}
\label{NONLINEAR}

The simulated data used to test FITEX for BAO extraction under more
realistic conditions are provided by the $\rm{\Lambda CDM}$ version of
the Hubble Volume Simulations conducted by the Virgo Consortium
(Evrard et al. 2002). One billion dark matter particles were simulated
with $h=0.7$, $\Omega_{\rm m}=0.3$, $\Omega_\Lambda=0.7$, $\Omega_{\rm
b}=0.04$, $\Omega_{\rm \nu}=0$, and the normalization of the matter power spectrum,
$\sigma_8=0.9$, in a cube with a comoving length of $L\,=\,3000$
$h^{-1}\,\rm{Mpc}$ and a mass of $2.25\cdot 10^{12}\,h^{-1}\,\cal{M}_\odot$ per
particle. The simulations were started with a `glass-like'
load (see e.g. Baugh et al. 1995). During the simulation itself, long range gravitational forces
were computed on a $1024^3$ grid yielding a Nyquist critical frequency
of $k_c = 1.07\; h\,\textrm{Mpc}^{-1}$. The short range gravitational
forces were computed via direct summation and softened on a scale of
$0.1\;h^{-1}\,\textrm{Mpc}$. We used the $10\times10\;\textrm{deg}^2$
fraction of the XW extended deep wedge which uses periodic boundary
conditions from a redshift of  $z=4.4$ to provide a survey lightcone up to the redshift $z_{\rm
max}=6.8$. The lightcone output comprises of data that includes cluster 
evolution and thus mimics real life observations. Larger distance from the observer correspond to higher redshifts where structures are less pronounced due to linear and non-linear growth.

\begin{table}[htbp]
    \caption[Power spectrum cubes]{Parameters of the cubes for the
    power spectrum estimation. Col.\,1: Cube number. Col.\,2: Comoving
    length of the cube (concordance cosmology). Col.\,3: Redshift
    range covered by the cube. Col.\,4: Fundamental mode of the
    discrete Fourier transformation which corresponds to the sample
    bin size of the power spectrum in $k$ space. Col.\,5: Nyquist
    critical wavenumber of the Fast Fourier Transformation.} \center
    \label{table cubedata} \begin{tabular}{lcccc} \hline \hline \\
    Sample & length & $z$-range & $\Delta k$ & $k_c$ \\ & \small{
    $[h^{-1}\,\rm{Mpc}]$} & & \small{ $[h\,\rm{Mpc}^{-1}]$} & \small{
    $[h\,\rm{Mpc}^{-1}]$} \\ \hline \\
        cube1 & 200 & 0.58 -- 0.68 & 0.031 & 4.0\\
        cube2 & 226 & 0.68 -- 0.79 & 0.028 & 3.6\\
        cube3 & 256 & 0.79 -- 0.92 & 0.025 & 3.1\\
        cube4 & 291 & 0.92 -- 1.1 & 0.022 & 2.8\\
        cube5 & 329 & 1.1 -- 1.3 & 0.019 & 2.4\\
        cube6 & 373 & 1.3 -- 1.6 & 0.017 & 2.2\\
        cube7 & 423 & 1.6 -- 2.0 & 0.015 & 1.9\\
        cube8 & 479 & 2.0 -- 2.5 & 0.013 & 1.7\\
        cube9 & 500 & 2.5 -- 3.2 & 0.013 & 1.6\\
        cube10 & 500 & 3.2 -- 4.1 & 0.013 & 1.6\\
        cube11 & 500 & 4.1 -- 5.4 & 0.013 & 1.6\\
        \hline
    \end{tabular}
\end{table}

\begin{figure}[htbp]
    \centering \includegraphics[width=8cm]{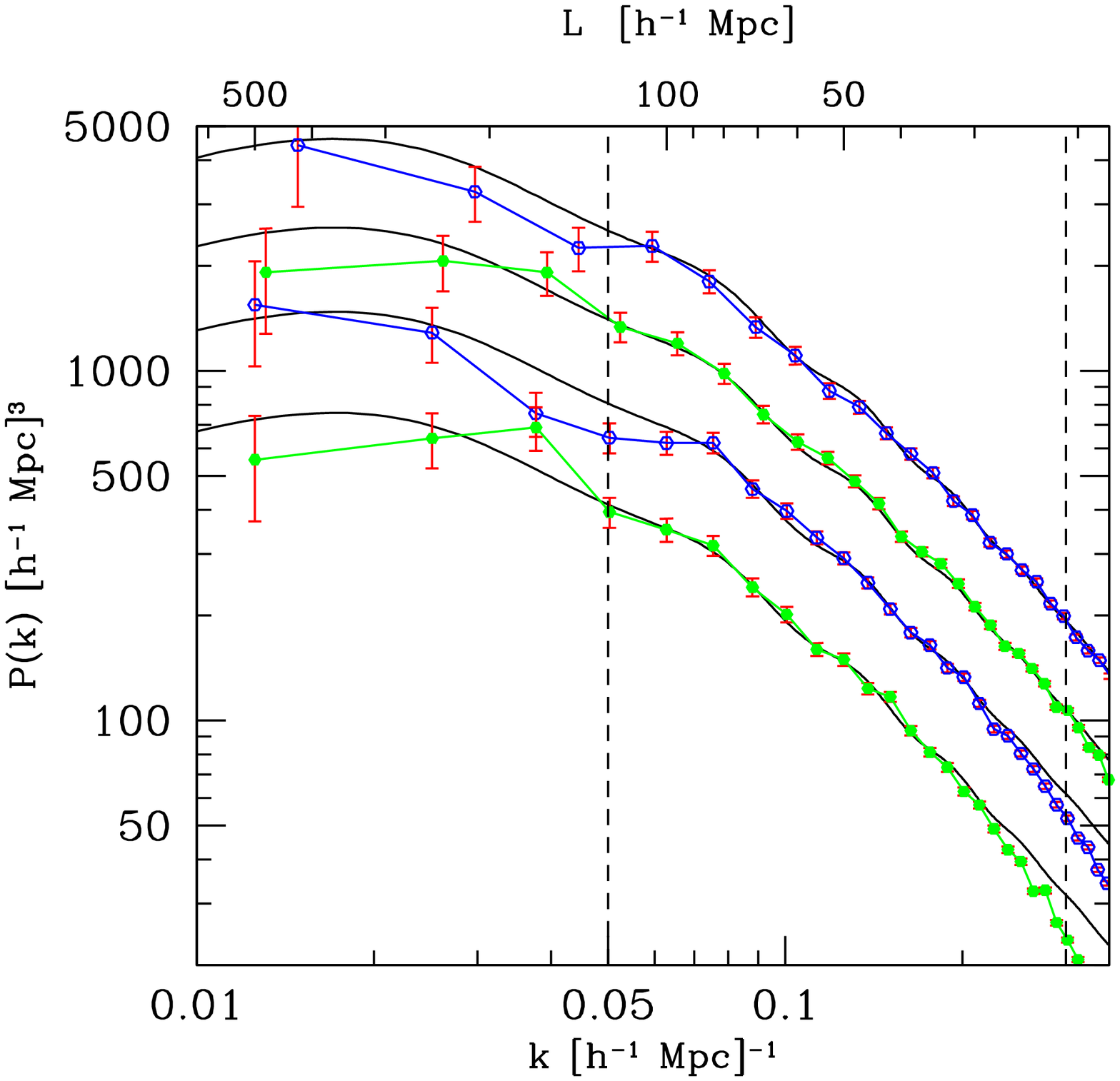}
    \centering \includegraphics[width=8cm]{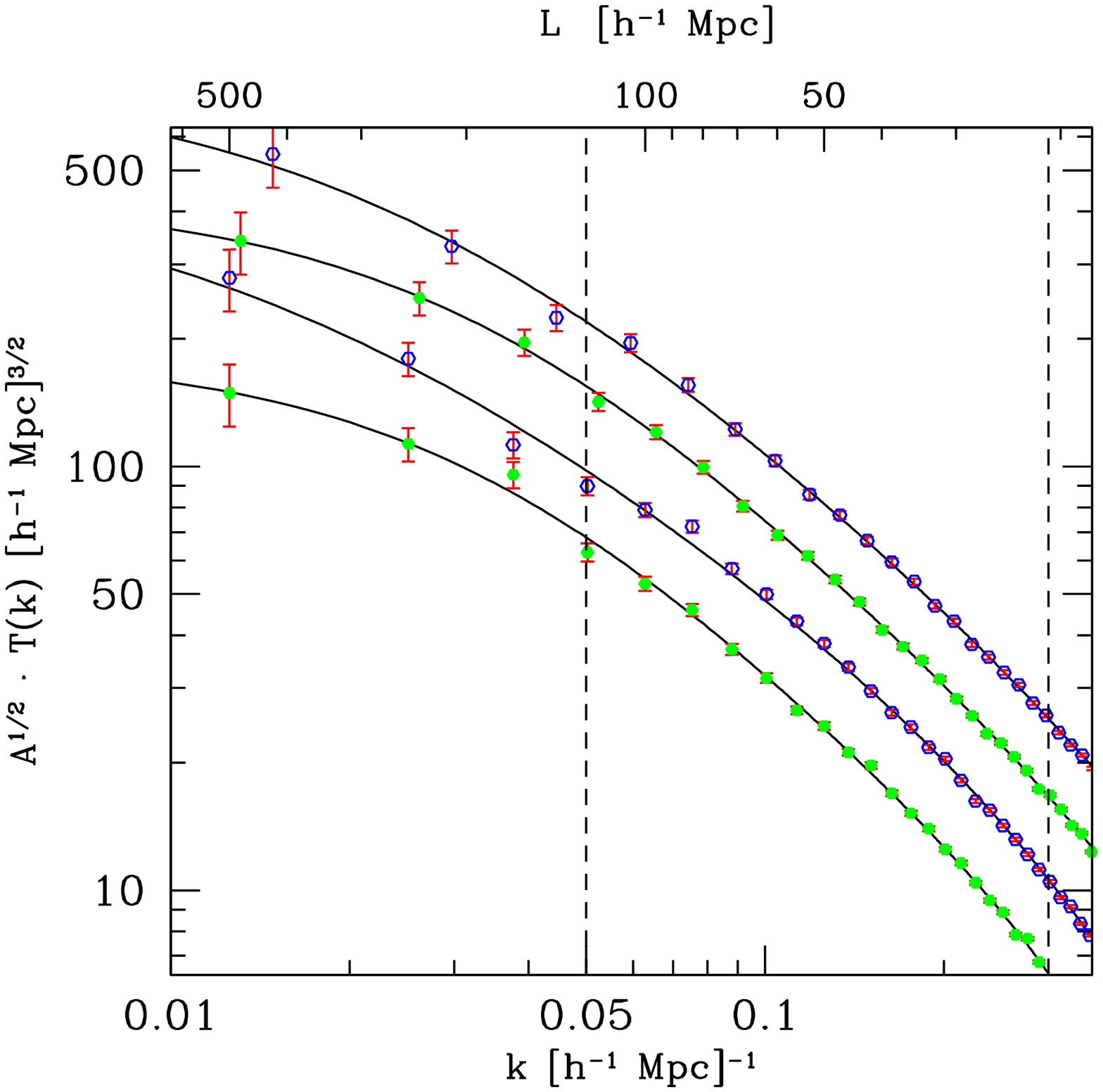}
    \caption[matter power spectra and transfer functions]{\small {\bf
    Upper panel:} Power spectra of the matter distribution from the
    cubes 7 (upper spectrum) to 10 (lower spectrum) of the XW deep
    wedge N-body simulation with superposed linear theory
    predictions. The linear matter power spectrum was taken from
    Eisenstein \& Hu (1998). Amplitude corrections for linear
    structure growth were implemented in the standard manner. For
    better illustration, the 2nd, 3rd and 4th spectrum is shifted
    downwards by the factors $\rm{dex}(0.1)$, $\rm{dex}(0.2)$, $\rm{dex}(0.3)$,
    respectively. The extraction of BAOs described in the main text is
    restricted to the $k$-range bordered by the two vertical dashed
    lines. {\bf Lower panel:} Same as upper panel for the
    corresponding boosted transfer functions with superposed fits of
    the phenomenological fitting functions. The fit at $z = 2.3$ uses the
    following parameters: $A =540.0$, $B =35.45 $, $k_1 =-0.32354$, 
    $\alpha =2.493$, $\delta =1.420$.}  \label{fig hubbleresults}
\end{figure}

Power spectra, $P_{\rm obs}(k,z,b)$, were estimated at different
redshifts and for different galaxy biasing schemes $b$ (see below)
with the Fast Fourier Transform using the variance-optimized method of
Feldman, Kaiser \& Peacock (1994) for cubes along the line-of-sight
(LOS) of the XW wedge. All the cubes were selected to fully fit into the lightcone in order to not introduce a non-trivial window function. The parameters of the cubes are given in
Tab.\,\ref{table cubedata}. Note that the spectroscopic SDSS survey of
normal galaxies covers an effective volume which is comparable to
cubes with $L\approx 500\,h^{-1}\,{\rm Mpc}$. The errors of the power
spectral densities at wavenumber $k$ were estimated with standard mode
counting arguments by
$\sigma_P/P(k)=\sqrt{2\pi}\{1+1/[P(k)\bar{n}]\}/kL$ to provide a first estimate.
 Here, $\bar{n}$ is the mean comoving particle number density in the box, and $L$ the
length of the box. We found that the initial glass load reduces the
shotnoise level significantly below the standard Poisson case at
redshifts $z>2$ (see Smith et al. 2003). Instead of correcting our
spectral estimator for this artifact (see, e.g., Smith et al.
2003), we could in the end neglect its impact on the cosmological
test, because our phenomenological fitting function was flexible
enough to compensate for the small residuals of this artifact seen in
our standard scale range $k<0.3\,h\,{\rm Mpc}^{-1}$.

In order to extract the BAOs from the power spectra $P_{\rm
obs}(k,z,b)$, we computed the corresponding boosted transfer
functions first, $\sqrt{A(k,z,b)}\,T(k)=\sqrt{P_{\rm obs}(k,z,b)/k}$,
assuming a primordial spectrum with $n\,=\,1$ (see
below). Fig.\,\ref{fig hubbleresults} shows the estimated power
spectra and the corresponding boosted transfer functions for an
unbiased dark matter particle distribution in configuration
space. Error bars include shot noise and sample variance. In most
spectra the 2nd and 3rd BAOs are clearly visible. For
$k\ge0.05\,h\,{\rm Mpc}^{-1}$, the phenomenological fitting function
(Eq.\,\ref{eq fit}) can be computed very accurately as it is
constrained by a large number of spectral data with relatively small
errors, that can be traced reasonably well by a power law. Sample
variance is negligible, as in this $k$-range, each spectral point
represents the average over a large number of modes. For
$k<0.05\,h\,{\rm Mpc}^{-1}$, only a very small number of spectral
densities, each estimated with a small number of independent modes,
exists and the phenomenological function has to fit a wiggle that
coincides with the turnover. Therefore, the results in the range
$0<k<0.05\,h\,{\rm Mpc}^{-1}$ vary quite a bit. However, we already
excluded this range from the cosmological test (see vertical dashed
lines in Fig.\,\ref{fig hubbleresults} and the discussion at the end
of Sect.\,\ref{EXTRACT}).

The extracted BAOs for the four cubes 7-10 of the XW deep wedge
simulation corresponding to the redshifts $z=1.8$-3.7 are shown in
Fig.\,\ref{fig realwiggles}.  Spectral densities in the range
$0.05<k<0.3\,h\,{\rm Mpc}^{-1}$ which is relevant for unbiased
cosmological tests are plotted (dots) together with the theoretical
wiggle function (Eqs. \ref{eq wiggles},\ref{eq correction}) multiplied
by an appropriate amplitude $\sqrt{A}$ (solid line). We see that even
for very large galaxy samples filling a box-like survey volume with a
length of $L=500\,h^{-1}\,{\rm Mpc}$ error bars are mainly determined
by sample variance and are of the same order as the amplitudes of the
BAOs. Therefore, only surveys with significantly larger survey volumes
can show the BAOs with higher significance.
 
\begin{figure}[htbp]
    \centering \includegraphics[width=4.3cm]{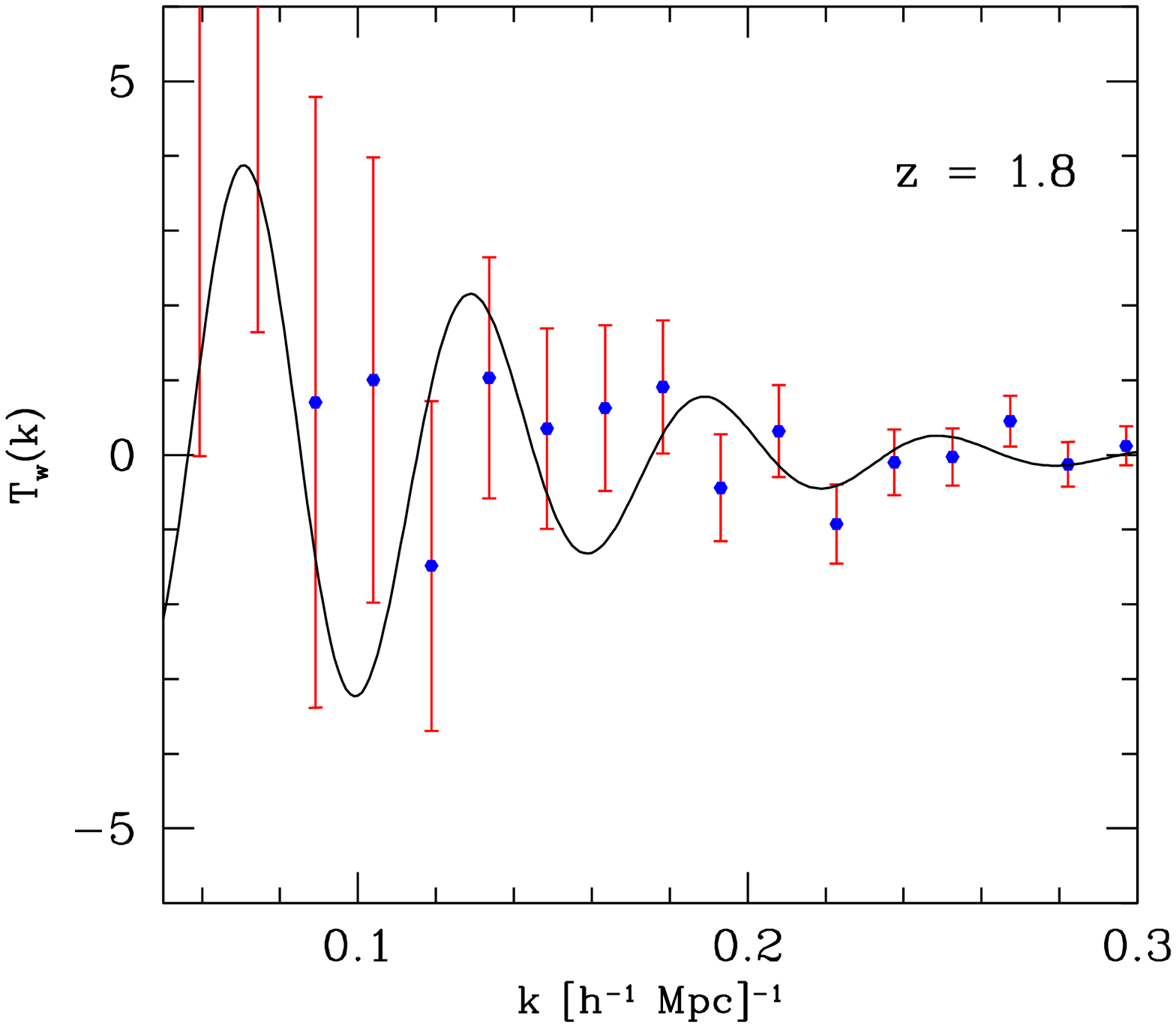}
    \centering \includegraphics[width=4.3cm]{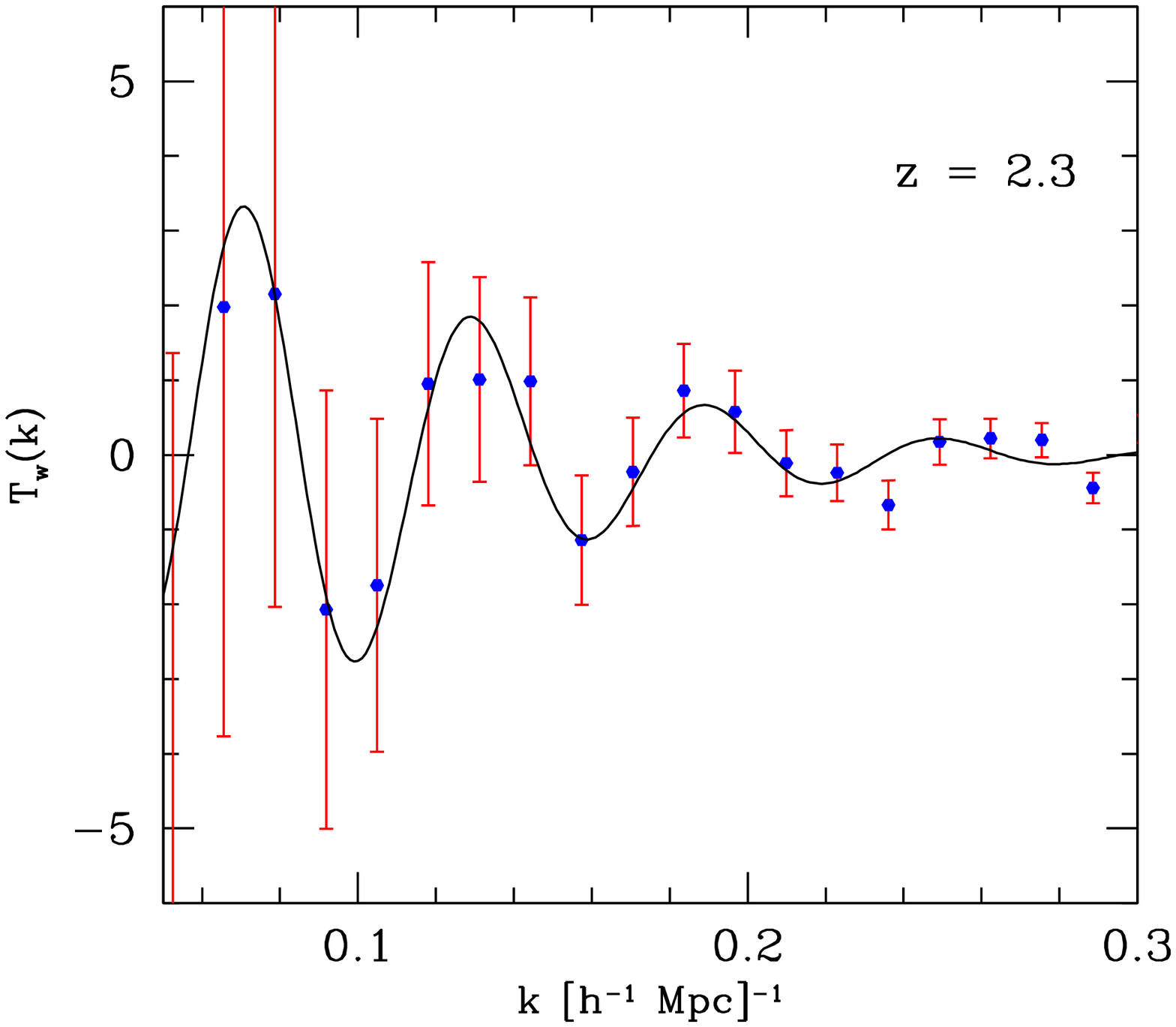}
    \centering \includegraphics[width=4.3cm]{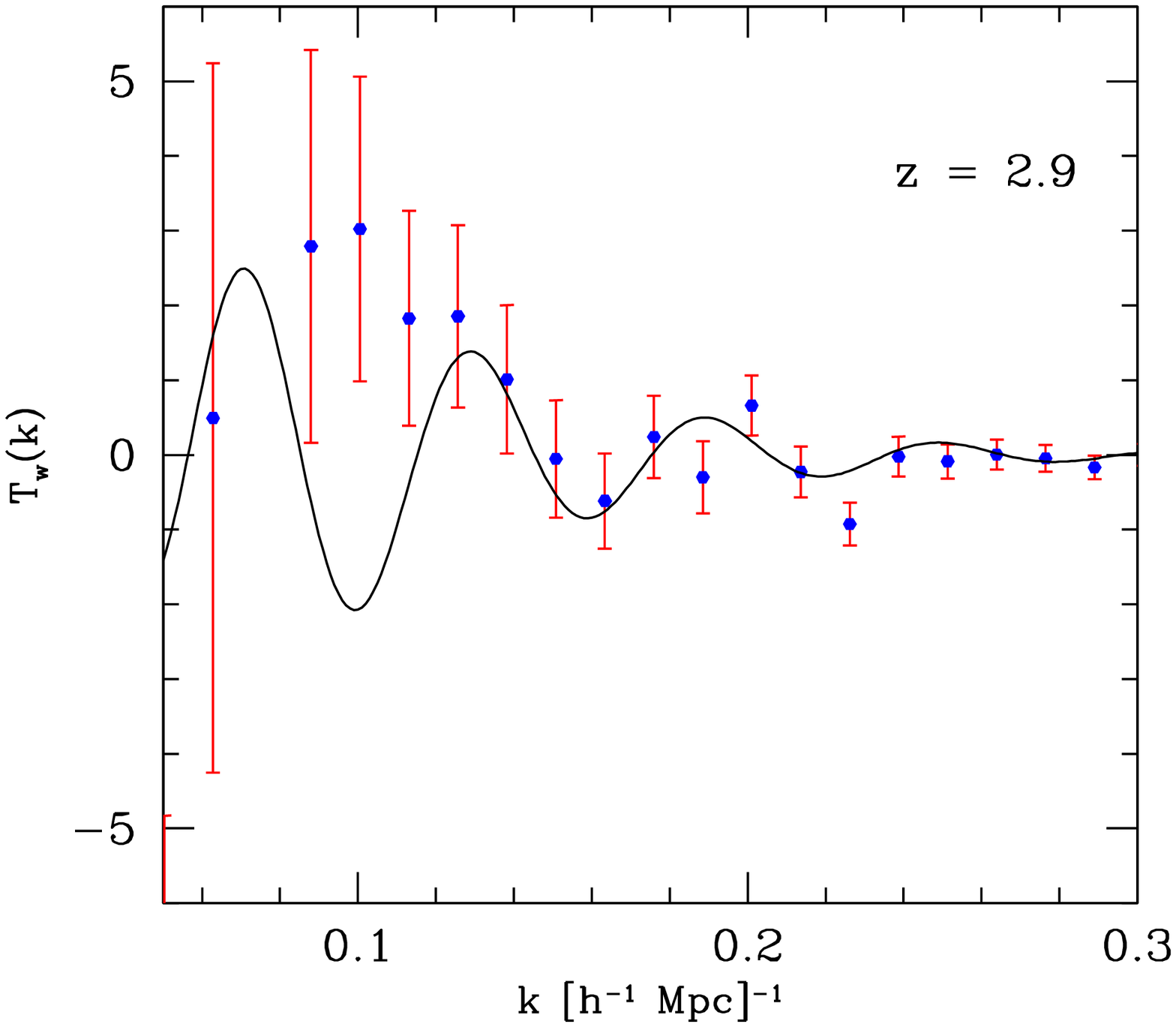}
    \centering \includegraphics[width=4.3cm]{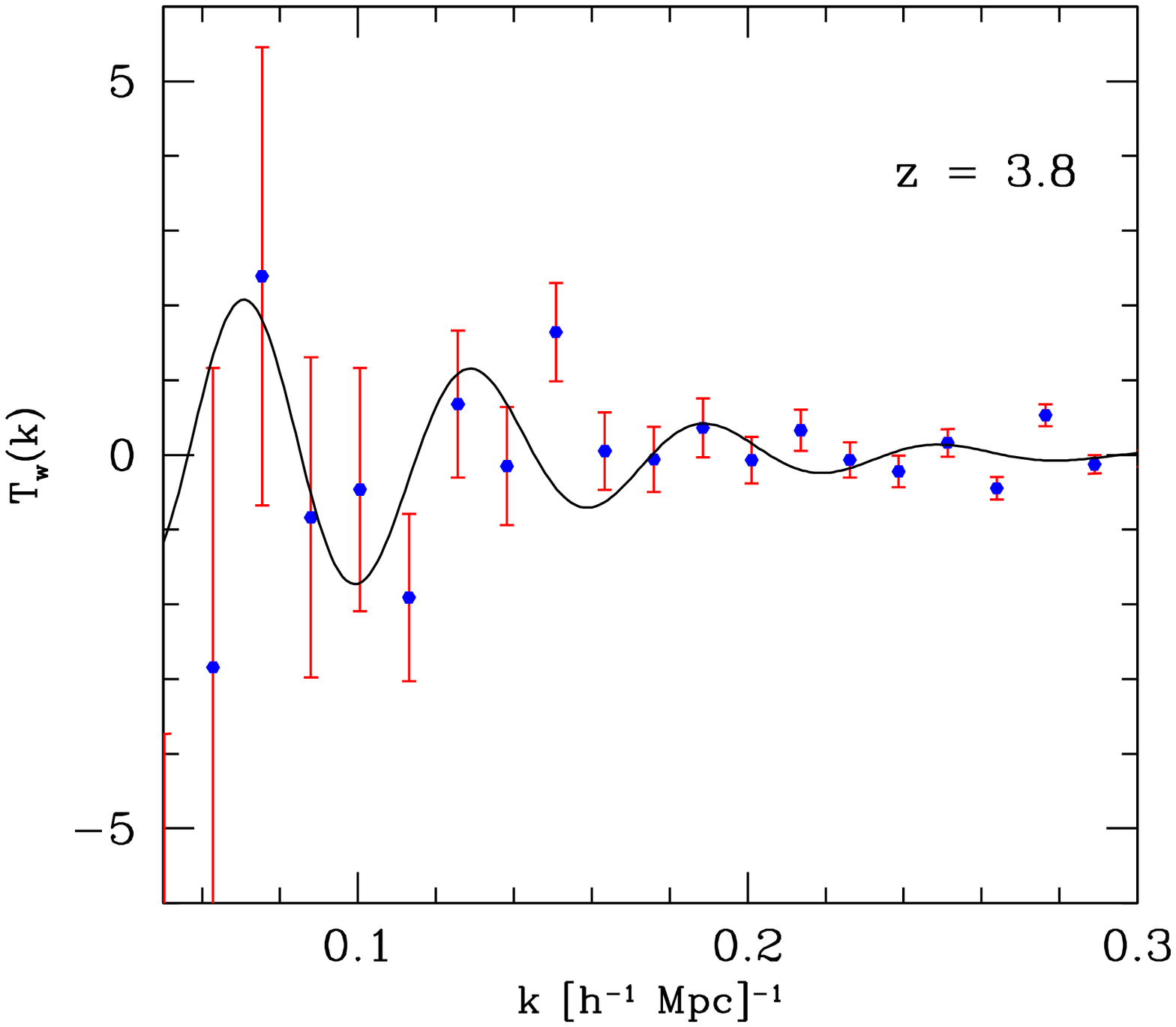}
    \caption[Wiggles in real space]{\small Extracted baryonic
    oscillations of dark matter particles (dots) with error bars
    including shot noise and sample variance at different redshifts
    ranging from $z\,=\,1.8$ (upper left panel) to $z\,=\,3.8$ (lower
    right panel). Superimposed is the theoretical wiggle function
    (Eq.\,\ref{eq wiggles}) with an appropriate amplitude.}  \label{fig
    realwiggles}
\end{figure}

\begin{figure}[htbp]
    \centering \includegraphics[width=9cm]{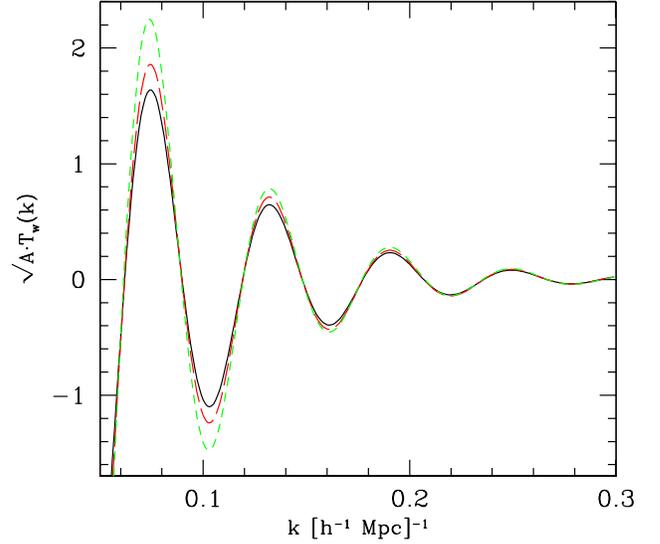}
    \caption[Extraction of different primordial power spectra]{\small Wiggle extraction 
from power spectra with different primordial $k$ dependencies.
The resulting oscillations using FITDEX on primordial models with a spectral index $n=1$ 
(solid line) is compared to a model with different spectral index $n=0.9$ (long dashed line) 
and a running spectral index with $\alpha=0.1$ (short dashed line).}
\label{fig running}
\end{figure}

We further see that the amplitudes of the wiggle functions in
Fig.\,\ref{fig realwiggles} decrease with increasing redshift, due to
growth suppression (we verified the amplitude change with $z$
according to standard linear structure growth). This is a direct
consequence of the fact, that distortions of the power spectrum are
directly translated into the boosted transfer function and thus into
distortions of the amplitude of the BAOs. To see how this works and how deviations from
the assumed primordial spectrum with $n=1$ affect the wiggle function,
we rewrite the primordial spectrum in the form $k^{1+2\epsilon}$ with
$|\epsilon|$ small compared to unity. For the assumed power spectrum,
$P(k,z,b)=A(k,z,b)k^nT^2(k)$, this choice gives
\begin{equation}\label{EX1}
\sqrt{\frac{P(k,z,b)}{k}}-F(k,b,z)=\sqrt{A(k,z,b)}k^\epsilon
\frac{\Omega_{\rm b}}{\Omega_{\rm m}}T_{\rm w}(k)\,,
\end{equation}
where the phenomenological function (Eq.\,\ref{eq fit}) effectively
fits the amplitude and shape-dependencies (see
Eqs.\,\ref{TK},\,\ref{T_b})
\begin{equation}\label{EX2}
F(k,z,b)=\sqrt{A(k,b,z)}k^\epsilon
\left[\frac{\Omega_{\rm c}}{\Omega_{\rm m}}T_{\rm c}(k)+
\frac{\Omega_{\rm b}}{\Omega_{\rm m}}\tilde{T}_{\rm b}(k)\right]\,.
\end{equation}
It is seen that the phases of the wiggle functions remain unchanged by
the FITEX procedure. This result is shown in Fig. \ref{fig running} where the extracted oscillations
from power spectra with different spectral index models in the primordial power spectra are plotted. The oscillations were retrieved by calculating a power spectrum with a spectral index $n = 1$ (solid line), an
index of $n = 0.9$ (long dashed line) and a simple running spectral index model $k^{1-0.1\ln{k}}$ (short dashed line) and applying the FITDEX method. The parameters are chosen to conservatively cover the range of primordial power spectra not excluded by WMAP 3rd year data (see Bridges et al. 2006).

Only the amplitudes of the wiggle functions are
modulated by possibly scale-dependent factors. Although a lot of
interesting information about galaxy formation and cosmology is
contained in the amplitude factors of the wiggle function, we skip
this detail here and perform an amplitude rescaling,
$\sqrt{A(k,z,b)}k^\epsilon\Omega_{\rm b}/\Omega_{\rm m}\rightarrow
\sqrt{A(k,b,z)}$, with $A(k,z,b)$ being a basically arbitrary function after
rescaling, summarizing distortions like structure growth, galaxy
biasing, redshift space effects, deviations from the
Harrison-Zel'dovich case etc. Thus, with FITEX we are not deriving a
pure undistorted wiggle function, but a spectrum which is subject to
all the distortions that work on the power spectrum.

While the oscillations extracted from the Hubble Volume Simulations are consistent with the theoretical template, the error bars prevent a precise evaluation of the accuracy of FITEX
(all 4 cubes will be combined later, see Fig. \ref{fig combwiggles}).
To get a better estimate of non-linearities that might wash out wiggles at higher $k$ values,
we use a non-linear analytic form of the power spectrum. This was computed by Jeong \& Komatsu (2006) using
3rd-order perturbation theory, which is able to model non-linear N-body matter power spectra to better than
1\% at $z>2$ and $k<0.3\,h\,{\rm Mpc}^{-1}$.

\begin{figure}[htbp]
    \centering\hspace{-0.8cm}
    \includegraphics[width=9cm]{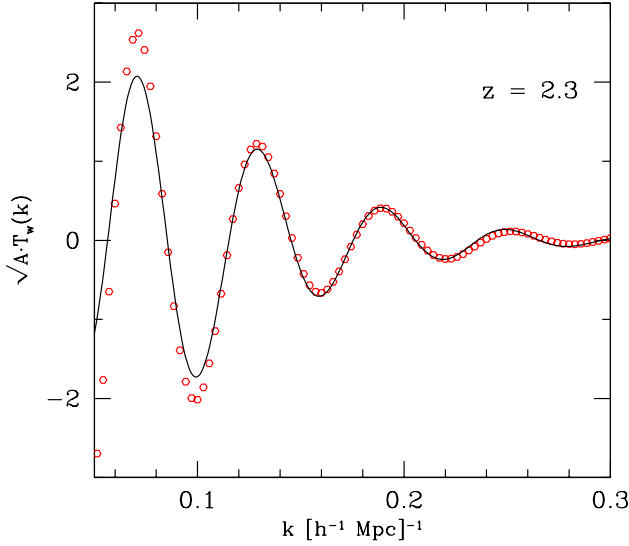}
    \caption[Extraction from analytical function]{\small 
    BAO extracted from an analytic power
    spectrum of Jeong \& Komatsu (2006) that includes non-linear growth at $z=2.3$ (dots).
    The continuous line is
    the theoretical wiggle function derived in Sect.\,\ref{BAOS},
    multiplied by an appropriate amplitude. See the main text for a
    more detailed discussion of the treatment of the amplitudes of the
    wiggle function and the analytic power spectrum.} \label{COMBWIG100}
\end{figure}

Figure \,\ref{COMBWIG100} shows the results of the FITEX extraction applied to the analytic non-linear
power spectrum using Hubble Volume parameters. As expected, the amplitude deviates from the theoretical prediction at small $k$-values, where the shape contribution of the baryons dominates (see. Fig. \ref{fig transfer}), but the phase of the oscillations in the analytical non-linear function is recovered to better than 1\% accuracy. This clearly shows, that, at least in theory, FITEX is able
to model non-linear effects and recover the oscillations with high accuracy.

\begin{figure}[htbp]
    \centering \includegraphics[width=9cm]{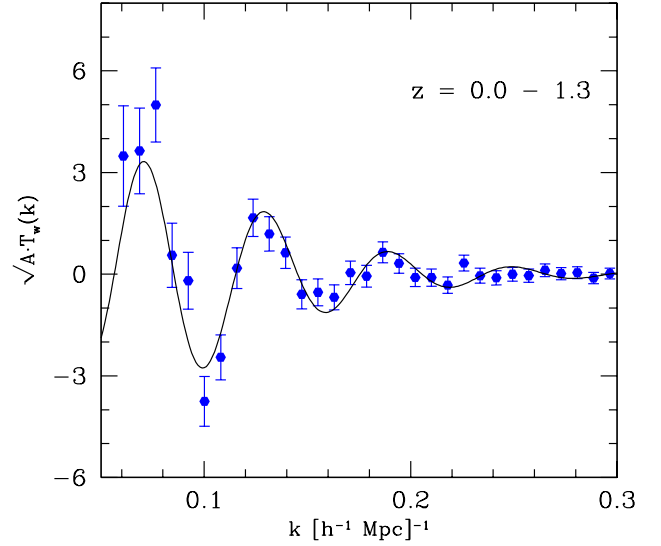}
    \caption[Oscillations at low redshift]{\small Extracted BAOs
    using the Hubble Volume Simulation covering one eighth of the sky
    from redshift 0 to 1.3. Error bars include sample variance and shotnoise but do not
    cover washing out effects introduced by the correlation of modes through non-linear growth and distant observer approximation.}
\label{fig octant}
\end{figure}

As an additional test, the power spectrum of the Hubble Volume Simulation (OCTANT data), a wedge covering on eighth of the sky from $z = 0 - 1.3$ including redshift-space effects, was computed with about 60 million particles in a fast Fourier box with a legnth of $1600\,h^{-1}\,\rm{Mpc}$ and the FITDEX algorithm used to extract the oscillations (Fig. \ref{fig octant}). The oscillations can be seen, although non-linear growth is heavily washing out the wiggles at $k>0.22\,h\,{\rm Mpc}^{-1}$.

Furthermore, we see a very good correspondence between
a FITEX-extracted wiggle function obtained from N-body simulations as well as the analytic non-linear 
power spectrum and the theoretical wiggle function of FITEX (Eq.\,\ref{eq wiggles}) in
the HETDEX redshift range. Both linear and non-linear structure growth
change the power spectrum and thus the BAOs, but FITEX still gives
quite stable and easy to analyze results for the 2nd to 5th BAO. The
presence of the BAOs is consistent with the results of recent
simulations of e.g. Springel et al. (2005) and Seo \& Eisenstein
(2005). However, no specific method to extract the BAOs was applied
and their simulations did not include lightcone effects. In the
following we concentrate on the effects of redshift space and
scale-dependent galaxy biasing on the results obtained with FITEX.

\section{Extracting BAOs in Redshift Space}\label{REDSHIFT}

The main problem in relating redshift space coordinates with comoving
coordinates is the differentiation between red-shifting due to the
expansion of the Universe and red-shifting due to peculiar velocities
of the measured particles. Two effects might be important. 

(1) In the linear regime and in combination with the distant observer
approximation, peculiar velocities introduce a boosting factor $\beta$
along the LOS. Only the $k$-modes parallel to the LOS,
$k_{\parallel}$, are boosted by the square of the factor
$1\,+\,\beta(z,b)k_{\parallel}^2/ (k_{\parallel}^2\,+\,k_{\perp}^2)$
with $\beta(z,b)=-d\ln D(z)/[d\ln(1\,+\,z)b(k,z)]$, $D(z)$ the linear
structure growth, and $b(k,z)$ the biasing parameter (Kaiser
1987). 

(2) In the non-linear regime, velocity dispersion of galaxies in
(partially) virialized structures cause the well-known `fingers of
God' effect in redshift space. We assume that the effect is mainly
restricted to small scales and is describable by a simple exponential
damping of the power spectrum by the factor $\exp(-\alpha
k^2_\parallel)$. Here, the redshift-dependent parameter $\alpha$ gives a
measure of the average pairwise velocity dispersion in the given
redshift shell.

\begin{figure}[htb]
    \centering \includegraphics[width=9cm]{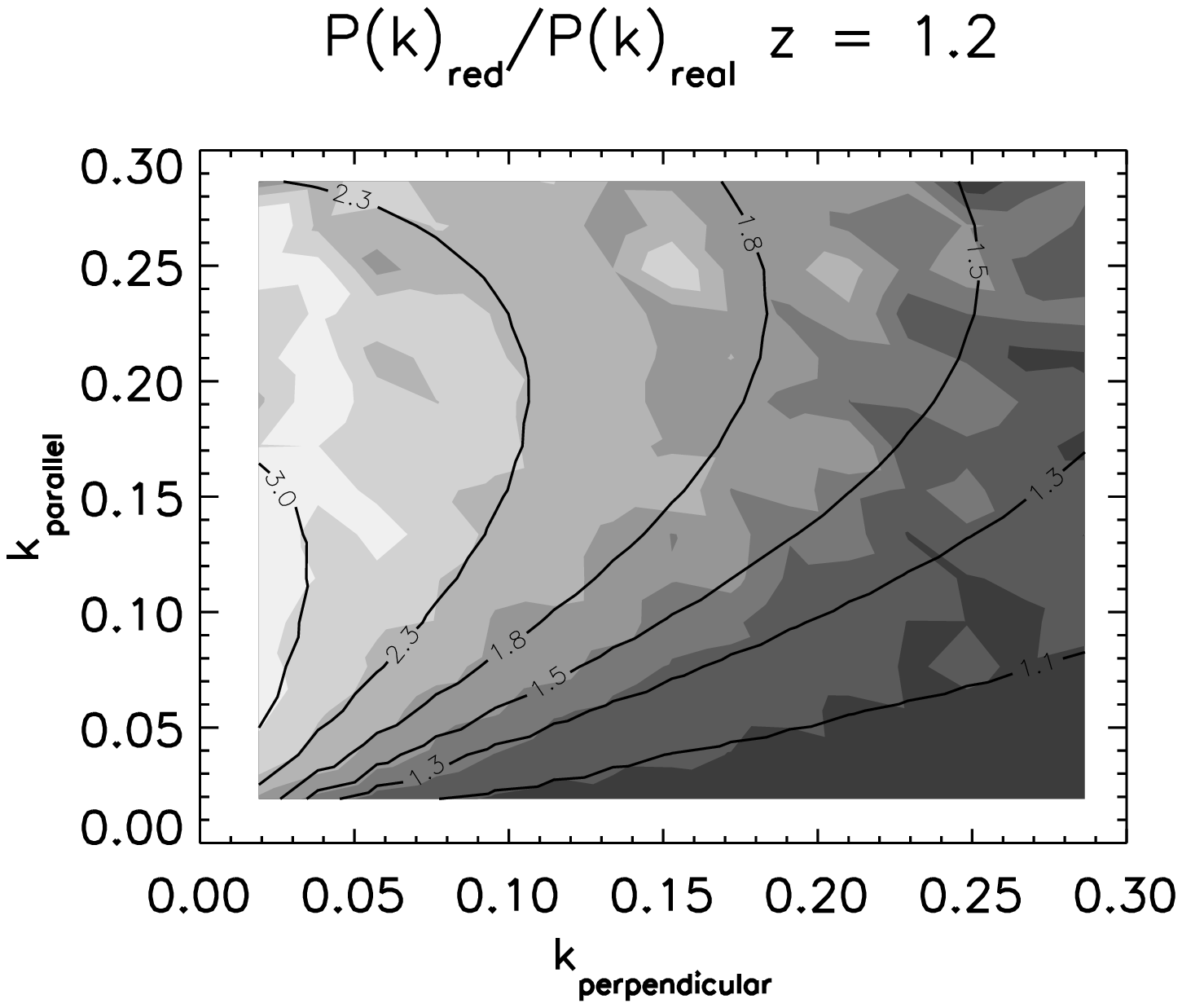}
    \centering \includegraphics[width=9cm]{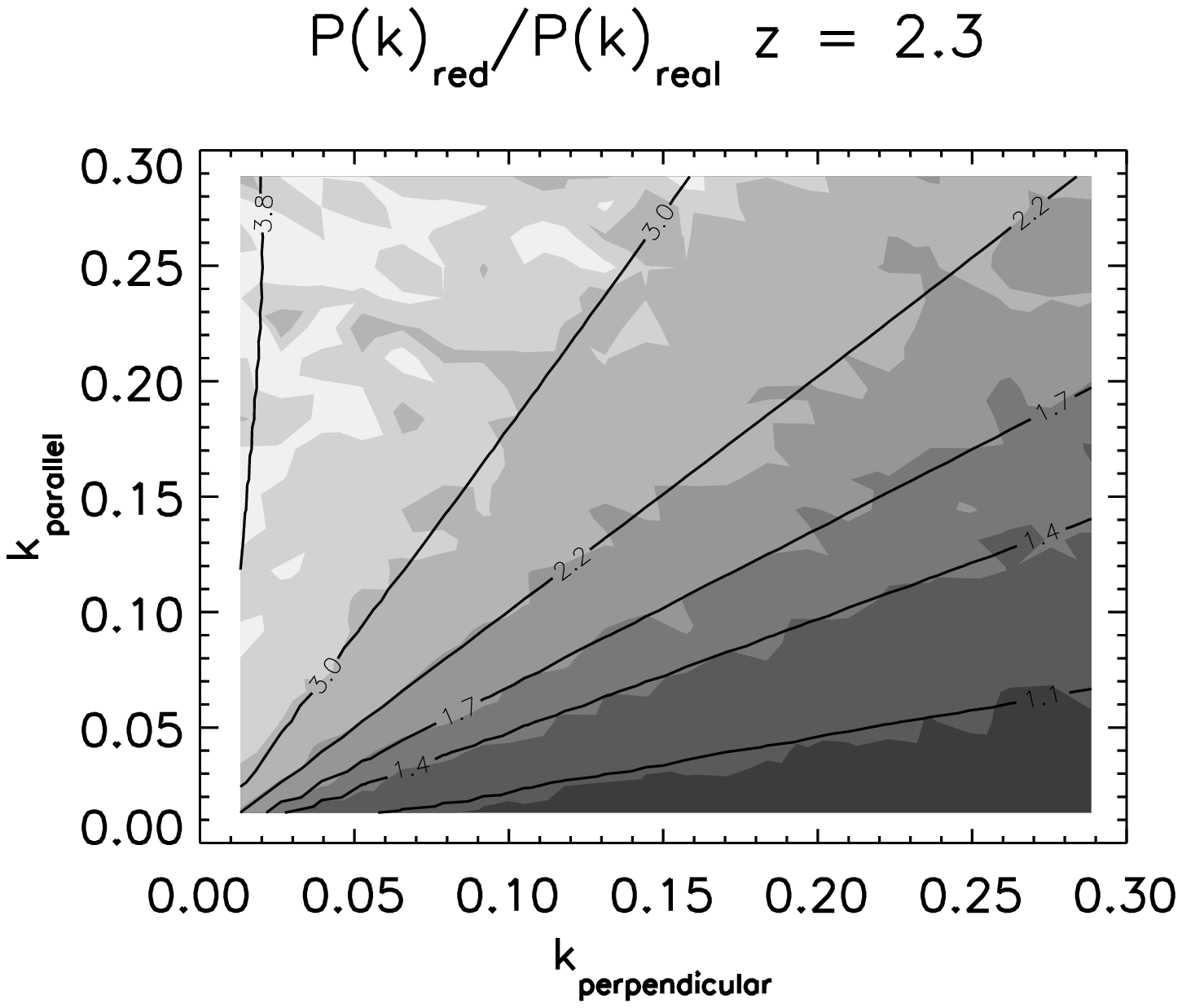}
    \caption[Redshift space effects]{\small Redshift space spectra
    divided by their corresponding real space spectra at different
    redshifts (filled areas). Overplotted lines represent the best fit of the kernel
    of Eq.\,(\ref{eq redcorrect}).}  \label{fig redshiftalpha}
\end{figure}

The combination of (1) and (2) constitute one version of the standard
dispersion model of peculiar velocities which relates the
one-dimensional redshift and real space power spectra by
\bq
    P_{\rm s}(k,z,b)\,=\,P(k,z,b)\,\int_0^1 d\mu
    \left[1\,+\,\beta(z,b)\mu^2\right]^2\,
    e^{-\alpha (k\mu)^2}\;,  \label{eq redcorrect}
\eq
where $\mu$ is the cosine between the direction of the actual
wavevector and the LOS, i.e., $\mu k = k_\parallel$, and
$k^2=k_\parallel^2+k_\perp^2$ (for more detailed models see
Scoccimarro 2004). On large scales, the real space power and thus the
wiggle function is boosted in a $k$-independent manner (Kaiser
factor). On small scales, the power spectrum is dampened in a
$k$-dependent manner by the factor $0.5\sqrt{\pi}\,{\rm erf}(\alpha
k)/\alpha k$, with erf the standard error function of statistics. In
Schuecker, Ott \& Seitter (1996) it is shown that Gaussian redshift
errors can be described with the same model (for
$\beta=0$). Eq.\,(\ref{eq redcorrect}) suggests that only the
amplitudes but not the phases of the BAOs can be changed by this type
of redshift space distortion.

\begin{figure}[htbp]
\centering \includegraphics[width=9cm]{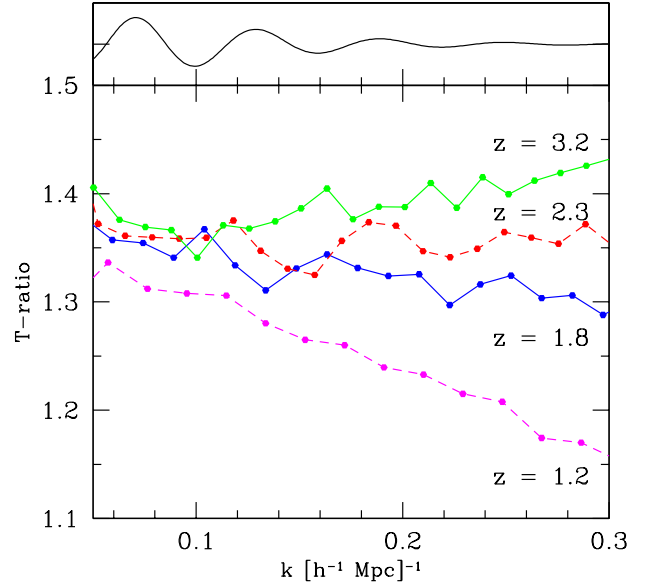}
\caption[Wiggles in redshift space]{\small Ratios of boosted 
redshift and real-space transfer functions for different redshifts. As
expected, the redshift space effects increase with decreasing redshift
in a smooth manner. Increasing ratios at high redshifts are an
artifact caused by the glass-like initial load of the N-body
simulations. To illustrate the relation between redshift space effects
and amplitudes of BAOs, the ratio between BAOs and the transfer function 
is plotted in the upper panel.}
\label{fig redwiggles}
\end{figure}

\begin{figure}[htbp]
\centering \includegraphics[width=8cm]{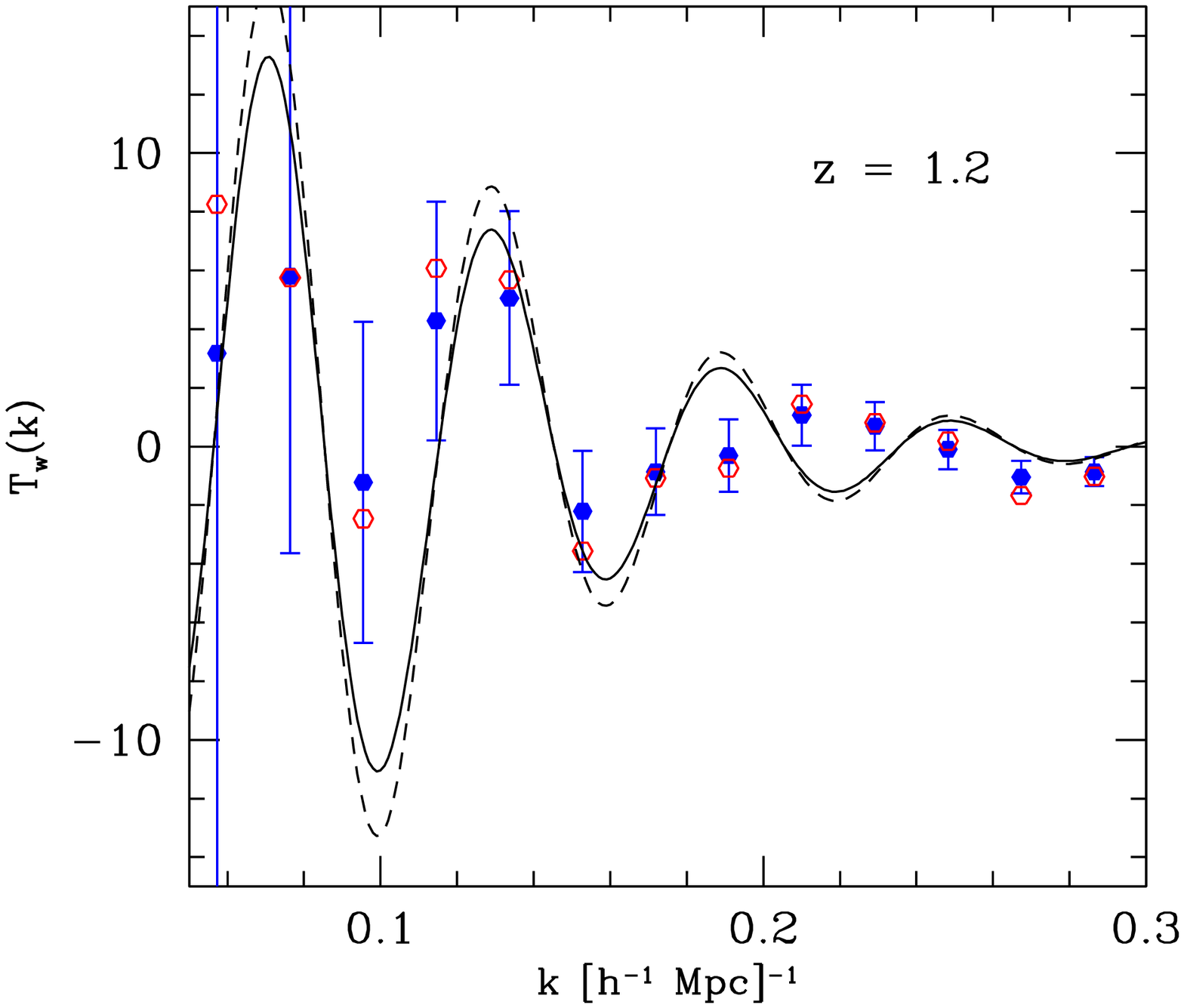}
\centering \includegraphics[width=8cm]{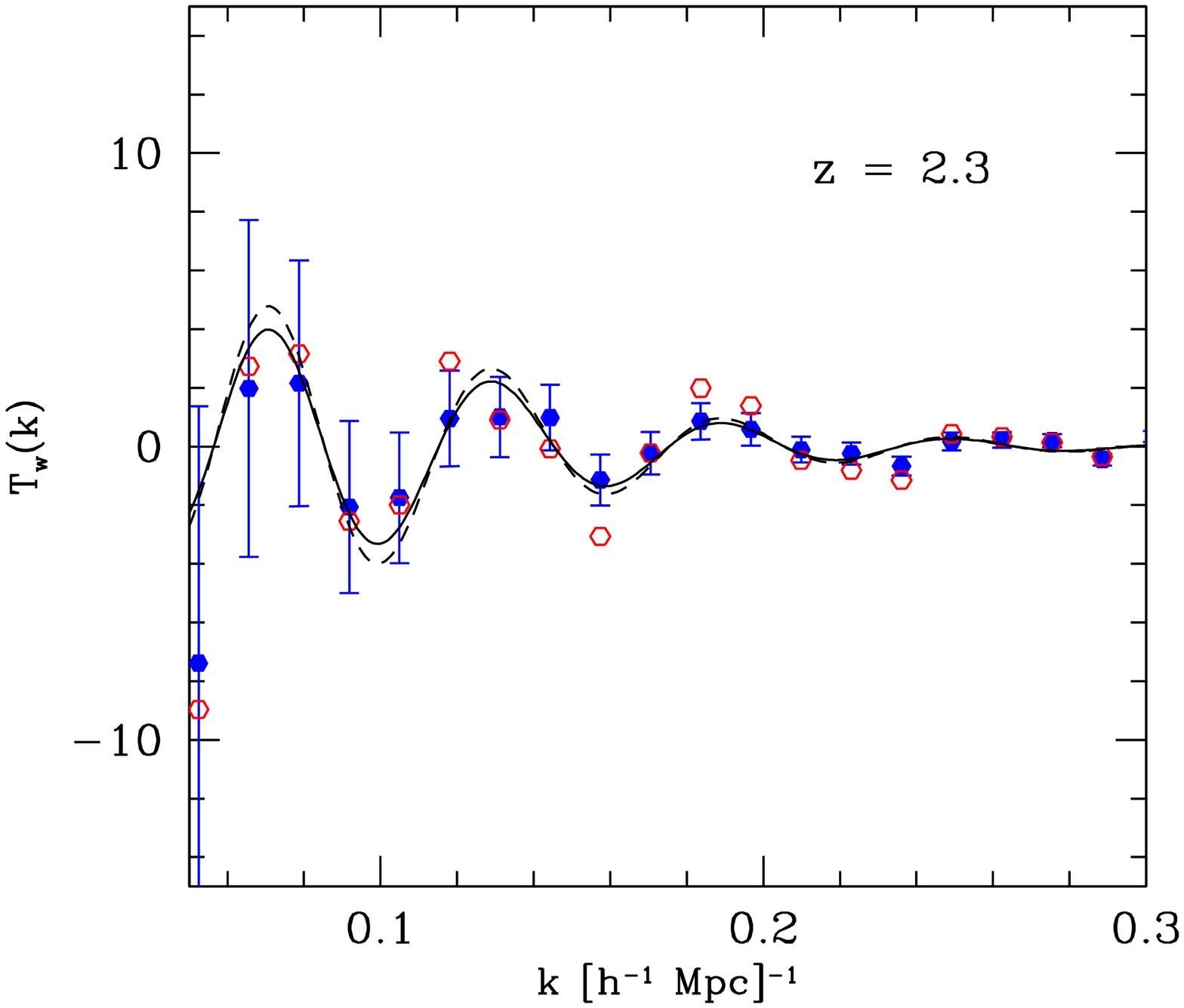}
\caption[Wiggles in redshift space]{\small  Comparison of wiggle
functions at $z=1.2$ (upper panel) and $z=2.3$ (lower
panel) obtained from real space (filled circles and continuous lines)
and redshift space (open circles and dashed lines). Note the overall
boost of the wiggle functions by redshift space effects which,
however, leaves the phases of the BAOs invariant. 
The non-linear growth effects, start to wash out the wiggles
 at $k > 0.2\,h\,\rm{Mpc^{-1}}$ in the upper panel.}  \label{RSBOOST}
\end{figure}

Figure \ref{fig redshiftalpha} illustrates the combined effect in more
detail. Two redshift space power spectra, as obtained from the Hubble
Volume Simulation, were divided by their corresponding real space
spectra at different redshifts and plotted as filled areas. The kernel
of Eq.\,(\ref{eq redcorrect}) was fitted to this ratio and the best
fit results superposed (black solid lines). We found that the model is
accurate on the 5\%-level over the complete redshift range. For
$z>2$, the parameter $\alpha$ is virtually zero because no
virialization had time to take place on the scales shown. As redshift
decreases, galaxy clusters start to form and virialize and the
$\alpha$ parameter starts to distort not the amplitude in the
two dimensional power spectrum depending on the angular component, like $\beta$ does, 
but changes the amplitude depending on the modulus of the $k$ vector.
This means that the shape of the wiggles could be
influenced by this effect as the amplitude of the power spectrum and
thus the amplitude of the wiggles is dampened at high $k$-values.

Fortunately, this anisotropy only gradually changes the amplitude of
each consecutive mode over which is averaged in a $k$-shell from
perpendicular to parallel modes. This is illustrated in Fig.\,\ref{fig
redwiggles} where the ratios of boosted transfer functions are
plotted. A rough idea about the relation between redshift space
effects and the amplitudes of BAOs is provided by the ratio between BAO and the
transfer function plotted in the upper part of Fig.\,\ref{fig redwiggles}.

Two examples of wiggle functions boosted by redshift-space effects and
extracted with FITEX are plotted in Fig.\,\ref{RSBOOST}. The
result of the anisotropic redshift space is that modes with a
more parallel component get more weight than those with a more
perpendicular component, which are not boosted. Whereas the overall
shape of the wiggle function appears to be slightly changed, the
phases of the wiggles are not distorted at all. Instead, the whole
wiggle function receives a boost of the monopole component of the
Kaiser factor.

The fitting function is obviously able to model the redshift space
distortions on the sub-percent level and oscillations can be derived. If
the wiggles were not extracted with FITEX-like methods but by dividing
the observed power spectrum by a standard non-oscillatory linear
theory power spectrum, as was done by in Springel et al. (2005,
Fig. 10) and Cole et al. (2005, Fig. 13), the baseline of the
oscillations would start to depart from zero at smaller scales. Such
an oscillation spectrum could not easily be used in a cosmological
test as it fit only nontrivial theoretical model spectra.

\section{Extracting BAOs from Biased Samples}\label{BIASING}

In the Hubble Volume, CDM particles are simulated. What is
actually observed are galaxies made out of dark matter and baryonic
matter, especially in the case of the proposed HETDEX survey, where
Lyman-$\alpha$ emitting galaxies are used as tracers of the underlying
matter distribution. Because of the different properties of dark
matter and baryonic matter (Benson et al. 2000, Le Delliou et al. 2006), both need not
necessarily follow exactly the same density distribution. 

To test the stability of FITEX under such conditions, we introduce some 
sort of biasing into the N-body simulations. Formally, biasing is described 
by the mean biasing function (Dekel \& Lahav 1999),
\bq
    b(\delta_d)\,\delta_d\;=\;\langle\delta_g\,|\,\delta_d\rangle\;=\;\int
    d\delta_g\,P(\delta_g\,|\,\delta_d)\,\delta_g\;,
\eq
that relates the dark matter distribution $\delta_d$ to the galaxy
distribution $\delta_g$ (Somerville et al. 2001). Here,
$P(\delta_g|\,\delta_d)$ is the local conditional biasing
distribution, i.e. the probability that a certain matter density
contrast $\delta_d$ with a variance $\sigma_d^2$ is converted into a
galaxy contrast $\delta_g$ with a variance $\sigma_g^2$. The biasing
function can also be expanded into the moments, $\hat{b}=\langle
b(\delta_d)\,\delta_d^2\rangle/\sigma_d^2$, and $\tilde{b}^2=\langle
b^2(\delta_d)\delta_d^2\rangle/\sigma_d^2$, with the biasing scatter
$\sigma_b^2=\langle(\delta_d-\langle\delta_g|\delta_d\rangle)^2
\rangle/\sigma^2_d$. Any local, non-linear and stochastic 
biasing relation can be described by these moments to second
order. While $\tilde{b}/\hat{b}$ is a measure for the non-linearity
and $\sigma_b/\hat{b}$ a measure of the scatter, the parameter
$\hat{b}$ characterizes the linear biasing such, that if $\hat{b} =
\tilde{b}$, then $\delta_g(\vec{x})=\hat{b}\delta_d(\vec{x})$ (Wild et
al. 2005).

\begin{figure}[htbp]
\centering 
\includegraphics[width=8.5cm]{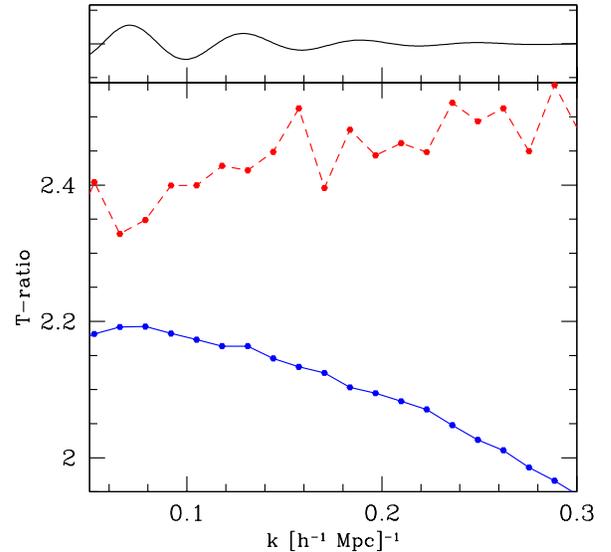}
\caption[Biased wiggles]{\small Ratios of biased
and unbiased  boosted transfer functions for two samples with different
scale-dependent biasing at $z=2.3$. A factor of 2 was subtracted from the strong bias (dashed line)
for enhanced visibility, while the weak biasing (solid line) remains unchanged. The theoretical wiggle
function is plotted on top of the figure to
illustrate the scale of the biasing effects.}
\label{fig_biaswiggles_a}
\end{figure}

\begin{figure}[htbp]
\centering 
\includegraphics[width=8.5cm]{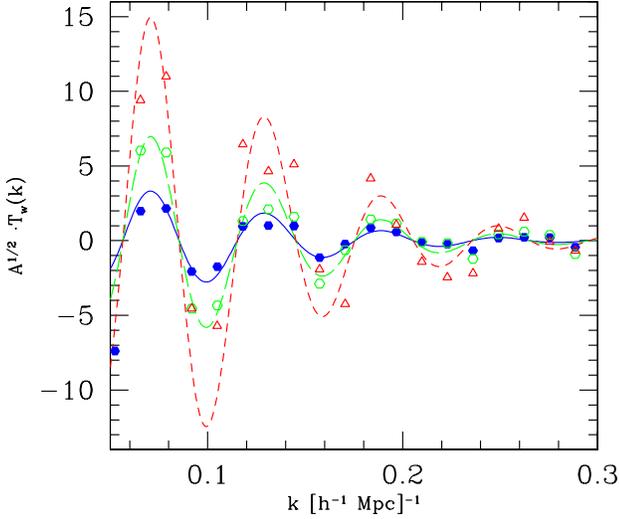}
\caption[Biased wiggles]{\small 
Boosted wiggle functions extracted with FITEX and theoretical wiggle
functions for the unbiased sample (filled circles and continuous
line), for the strongly biased sample (triangles and shot-dashed line)
and for the less biased sample (open circles and long-dashed
line). The theoretical wiggle functions are multiplied by the
$\hat{b}$-moment of the biasing function.}
\label{fig_biaswiggles_b}
\end{figure}

We use the simple stochastic and slightly
non-linear Eulerian biasing scheme  (e.g. Cole et al. 1998, Yoshida et al. 2001). 
First, local overdensities for each dark matter particle are computed
by smoothing the distribution with a Gaussian,
$\textrm{exp}(-r^2/2r_0^2)$ and $r_0 = 3\,h^{-1}\,\textrm{Mpc}$. The
overdensities $\delta_d$ are then transformed into normalized
overdensities, $\nu = \delta_d/\sigma_d$, with $\sigma_d$ being the root
mean square of the $\delta_d$-values.  Finally, a function is
introduced that describes the probability of a dark matter particle
with overdensity $\delta_d$ to be tagged as a `galaxy',
$P(\nu)\sim(\nu-\nu_c)^{\alpha}$, with $\nu_c$ being a threshold at
which the probability is set to zero. This simulates the lower
probability of galaxies forming in low density regions like voids. The
parameter $\alpha$ controls the stochastic spread of the biasing
function as well as its non-linearity. A value of $\alpha\,=\,0.2$ is
a good compromise between reasonable non-linearity and a comparably
tight spread. Two samples with $\nu_c\,=\, -0.2$ and 0.6 were
generated. Both parameters, $\alpha$ and $\nu_c$, were adjusted to
model the biasing parameters and number densities found by
observation. SUBARU data (Hamana et al. 2004) suggests a comoving LAE
density of $\bar{n}\,=\,0.00301\,\pm\,0.00194\,h^3\,\textrm{Mpc}^{-3}$
and a biasing parameter of $\hat{b}\,=$ 4-5 at a redshift $z\,=\,5$,
which should decrease slightly with redshift. As discussed in
Sect.\,\ref{NONLINEAR} the biasing function $b(\delta_d)$ should
translate into a boost of the power spectrum $b^2(k,z)$. Because the
average slope of the biasing function is $\hat{b}$, to first order
we expect $P_{\rm bias}(k)=P(k)b^2(k,z)\sim
P(k)\hat{b}^2(z)$.

The two prescriptions lead to a biasing that either decreases
or increases with $k$ (see 
Fig.\,\ref{fig_biaswiggles_a}). Both values are consistent with the
calculated $\hat{b}$ parameter. The comparatively strong scale dependence
of the biasing becomes evident when the ratios of
the boosted transfer functions are compared with the ratio of the
wiggle amplitude to the transfer function plotted on top of
Fig.\,\ref{fig_biaswiggles_a}.

Figure \ref{fig_biaswiggles_b} illustrates, that FITEX
extracts BAOs even from biased galaxy distributions. Plotted are the
wiggle functions of the less strongly biased sample (open circles),
the more strongly biased sample (open triangles) and the unbiased
sample (filled circles) at $z\,=\,2.3$. The unbiased and biased oscillations vary
because of shot noise, as the biased samples have a 2 and 10 times 
lower point density, respectively.
Superimposed are the theoretical wiggle
functions (Eq.\,\ref{eq wiggles}), normalized to the unbiased wiggle
function (continuous line) and multiplied by the first moment of the
biasing, $\hat{b}$ (dashed lines). The boost of the power spectrum
obviously translates into an amplitude boost of the BAOs. As in
redshift space, only the phenomenological fitting function as used in
FITEX extracts the BAOs with a zero baseline, unlike when using a
linear theory power spectrum.

However, not only the overall boost of the power spectrum, but also
the $k$-dependency of the biasing function should directly affect the
amplitude of the BAOs. We tried to verify this by fitting the $T_w(k)$
function supplemented by a `linear biasing model' to describe the
extracted wiggle function $T_{\rm ex}(k)=(a_0\,+\,a_1\,k)T_w(k)$. We
notice that both parameters $a_0$ and $a_1$ are highly degenerate,
mainly due to the fact that only data points at the extrema of the
wiggle function significantly contribute to the fit. Data points at or
near the nodes of the wiggle function are very insensitive to changes
of the amplitude. Due to the degeneracy, any amplitude
$a_0$ can easily be compensated by an appropriate value of the
parameter $a_1$.  Though the full fits had the expected
slopes, the effects were very small and of no significance for the
cosmological test. We may thus conclude that in a cosmological test,
which is mostly about detecting phase-shifts, it seems to be enough to
fit the theoretical wiggle function with only
an amplitude parameter, when the $k$-dependent biasing is as weak as
simulated here. However, more realistic simulations are needed to
analyze higher order effects.

\section{Cosmological Tests with BAOs}\label{CTEST}

The cosmological test compares the phases of the extracted BAOs with the theoretical
wiggle function, projected onto the hypersphere of the observer.
As the theoretical template does not model the amplitude of the oscillations with high
accuracy, we are not able to include amplitude effects in the cosmological test and marginalize
over the wiggle amplitude. Future tests should include this parameter, which is sensitive to $w$
by e.g. structure growth.

A general feature of the test is that one first has to assume a certain
reference cosmology to get the metric scale of the wiggle function
from the data, and to perform in a second step the comparison of the
metric scale of the theoretical wiggle function relative to the chosen
reference cosmology (Glazebrook \& Blake 2005). In the present paper, we are interested in
the performance of the BAOs extracted with FITEX in constraining the
redshift-independent part of the $w$ parameter of the dark energy. The
amplitude of the theoretical function is also fitted but later
marginalized. The remaining cosmological parameters are fixed to their 
values given at the end of Sect. \ref{INTRO}.

\begin{figure}[htbp]
    \centering \includegraphics[width=8.5cm]{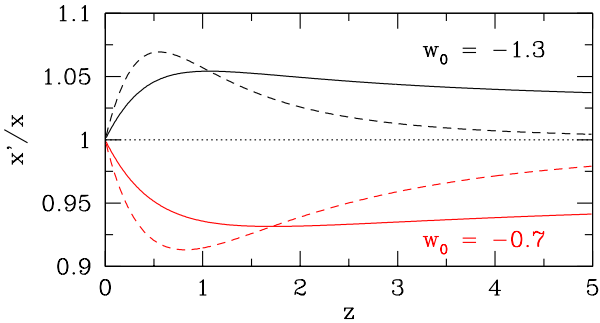}
    \caption[$w_0$ scaling] {\small Transformation of perpendicular
    (solid) and parallel (dashed) distances in comparison to the
    concordance model with $w_0\,=\,-0.7$ (upper part) and $w_0\,=\,-1.3$
    (lower part).}  \label{fig scales}
\end{figure}

As metric scales, comoving distances of the BAOs parallel and
perpendicular to the LOS are used. Recall that in the plane-parallel
or distant observer approximation, comoving distances between two
points with the same angular positions but different redshifts are
$x_{\parallel}=\int_{z_1}^{z_2}\frac{dx}{dz}dz=
\frac{c}{H_0}\int_{z_1}^{z_2}\frac{dz}{E(z)}$, where
$H(z)=H_0E(z)$ is the Hubble parameter at redshift $z$ and $E(z)$ the
$w$-dependent transformation of the Hubble constant, $H_0$, from
redshift zero to $z$ (e.g., Peebles 1993). To compute not distances
between two points in space with different redshifts, but parallel
scaling factors between the two sets of cosmologies denoted by primed
and unprimed symbols, one has to take the limit $\Delta
z=z_2-z_1\rightarrow0$, which yields
\begin{equation}\label{PARALLEL}
x'_\parallel\,=\,x_\parallel\,\frac{dx'}{dx}\,. 
\end{equation}
Comoving perpendicular separations represent distances two points with
the same redshift but different angular positions are separated,
$x_\perp=\theta\int_{0}^{z_2}\frac{dx}{dz}\,dz$, where $\theta$ is the
angle between the two points. The scaling relation for the perpendicular
component between two cosmologies thus becomes
\begin{equation}\label{PERP}
x'_{\perp}\,=\,x_\perp\,\frac{x'}{x}\,.
\end{equation}
In Fig. \ref{fig scales} the transformation ratio of the wiggle
function is plotted at different redshifts for two cosmologies that
differ from the reference model only in $w_0$ (the
redshift-independent part of $w$). The reference model is the
concordance cosmology. Fig.\,\ref{fig scales} shows that the largest
deviations from the reference model appear at redshifts in the range
$0.5<z<1.5$.  Differences in parallel and perpendicular scaling
introduce anisotropies that could be measured, too (see below).

\begin{figure}[htbp]
    \centering \includegraphics[width=8.5cm]{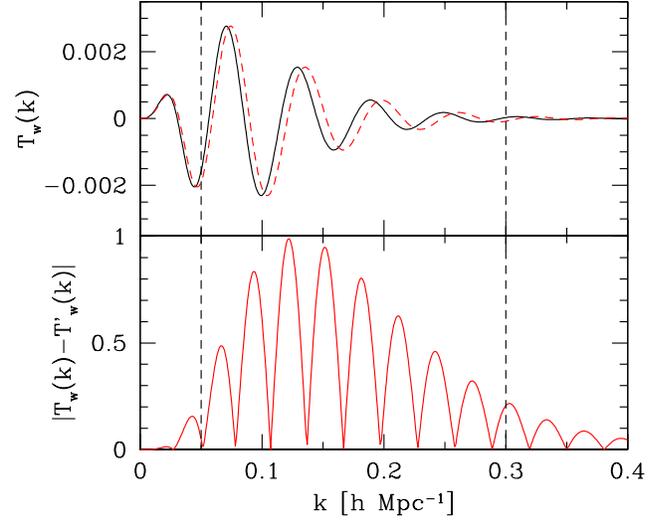}
    \caption[Scaled transfer functions] {\small Scaled wiggle
    functions at $z\,=\,2.5$ for $w_0=-0.7$ (dashed line) compared
    to the $w_0\,=\,-1$ cosmology (solid line, top panel). The normalized modulus of the 
    difference between the reference cosmology and the scaled cosmology is shown in the bottom panel.
    The dashed vertical lines mark the region relevant for the cosmological test.}
    \label{fig wigglescales}
\end{figure}

Figure\,\ref{fig wigglescales} shows the scaled wiggle function at
redshift $z=2.5$ for $w_0=-0.7$. The reference and the distorted wiggle 
function are plotted.
Note that an elongation of the coordinates in real space
leads to a compression of the coordinates in $k$-space and vice
versa. In the upper panel, it can be seen that the wiggle function
gets more and more out of phase as $k$ increases. Thus the sensitivity
of a cosmological test should increase at high values of $k$. However,
the amplitudes of the oscillations decrease exponentially with $k$
(e.g., Silk damping). This can be seen in the lower panel of
Fig. \ref{fig wigglescales} where the scaled wiggle function is
subtracted from the reference ($w_0=-1$) wiggle function: Though the
difference in phase is increasing, the amplitude decreases so fast
that the difference between the projected oscillations
increases at small values of $k$ and decreases at high values of $k$.

The present version of the cosmological test comprises the following
steps: (1) Angular and redshift coordinates of the `observed'
(simulated) data are transformed into comoving coordinates using a
concordance model as the reference cosmology. (2) The comoving data
sample is Fourier-transformed to estimate the boosted transfer
functions. (3) BAOs are extracted with FITEX. (4) The theoretical
wiggle function is computed for the corresponding cosmological
model. (5) The parallel and perpendicular scaling factors between the
test cosmology and the reference cosmology are computed at the
redshift of the data sample. (6) The theoretical wiggle function is
scaled by constructing a three-dimensional wiggle function, scaling
each mode and collapsing it to one dimension. (7) The observed wiggle
function is compared to the re-scaled theoretical wiggle function
assuming Gaussian statistical errors ($\chi^2$ test). Only the steps
(4-7) have to be repeated to test a different cosmology, allowing a
fast test of various parameters.

Note that the three-dimensional wiggle function is only constructed to
simulate the effects of different parallel and perpendicular scaling
factors. It does not address the anisotropy introduced by redshift
space effects, because we found that this only boosts the various
modes but not change the phases of the theoretical wiggle function
(see Sect \ref{REDSHIFT}). The boost effect of the redshift space
distortions, along with other boosting effects like biasing and growth
suppression, will force us to add an additional parameter to the
cosmological test.

\begin{figure}[htbp]
    \centering \includegraphics[width=4.3cm]{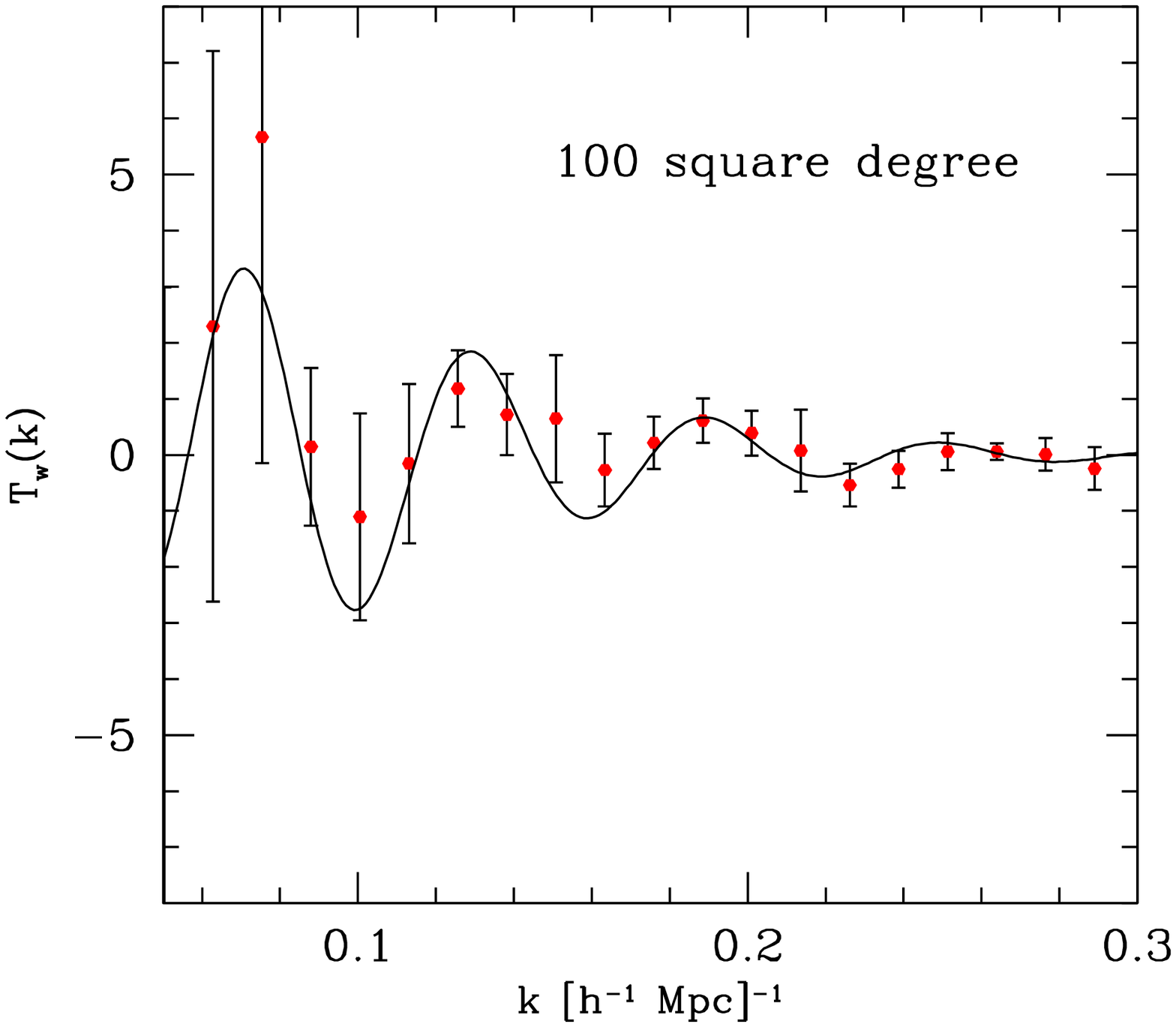}
    \centering \includegraphics[width=4.3cm]{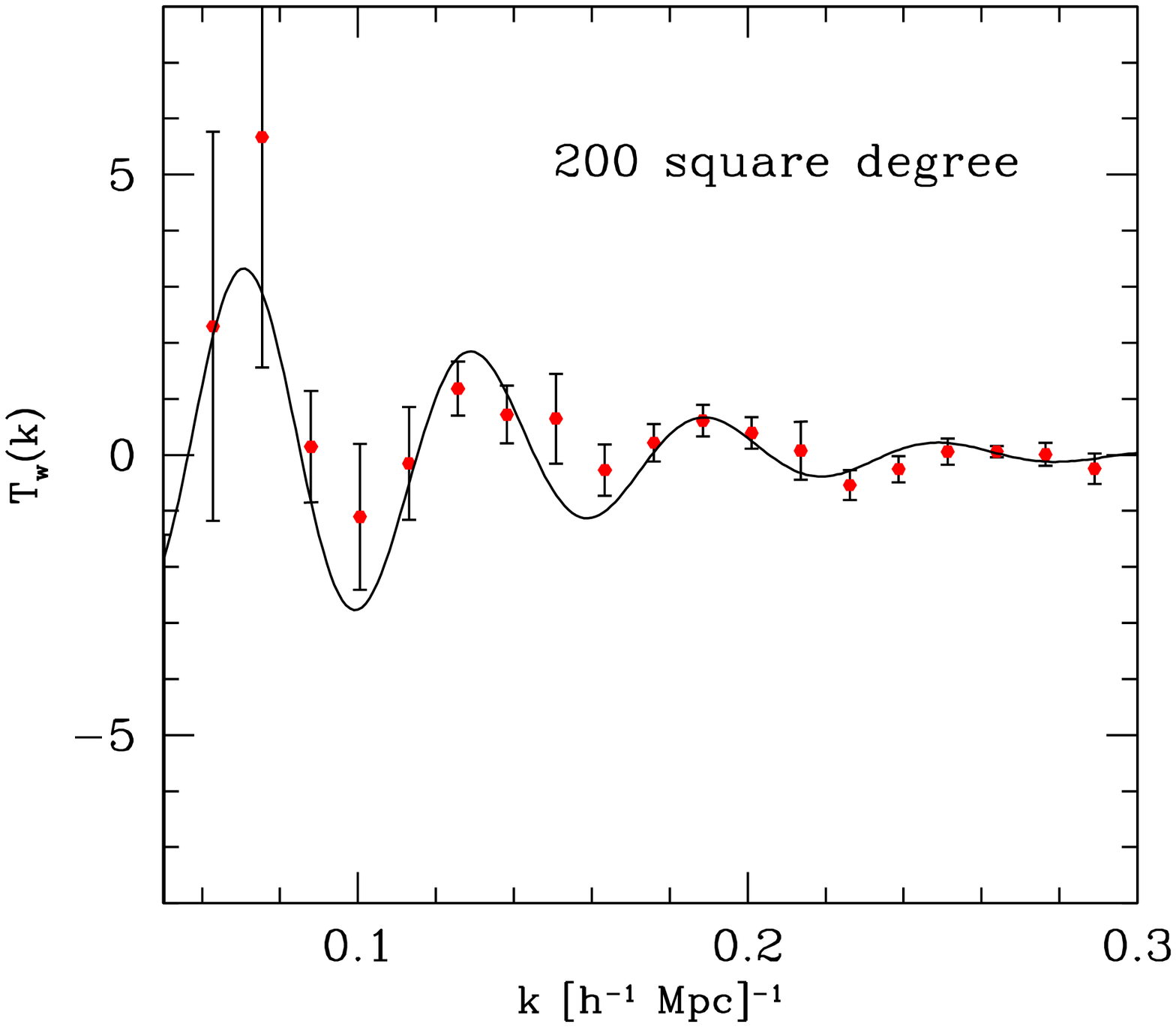}
    \centering \includegraphics[width=4.3cm]{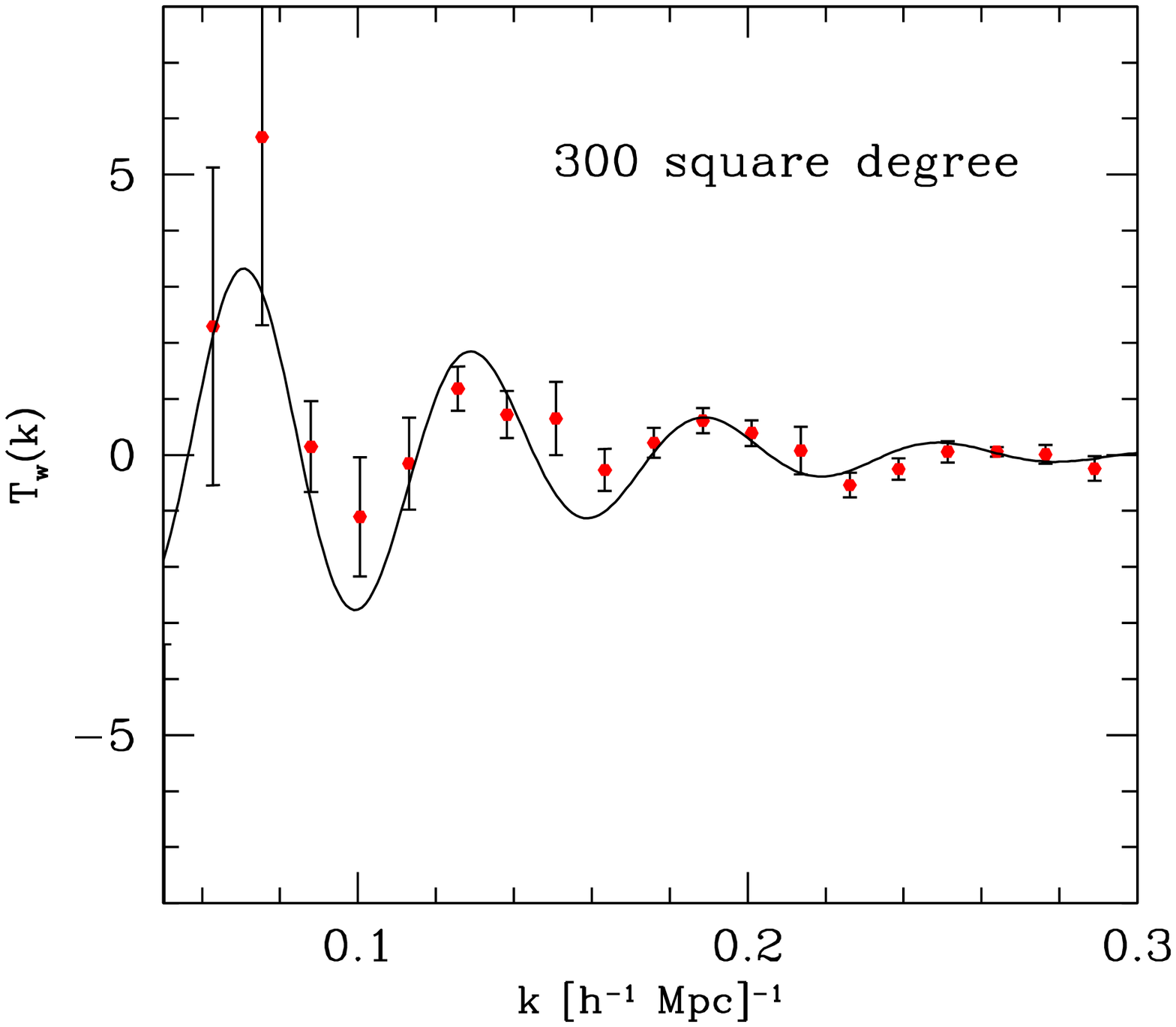}
    \centering \includegraphics[width=4.3cm]{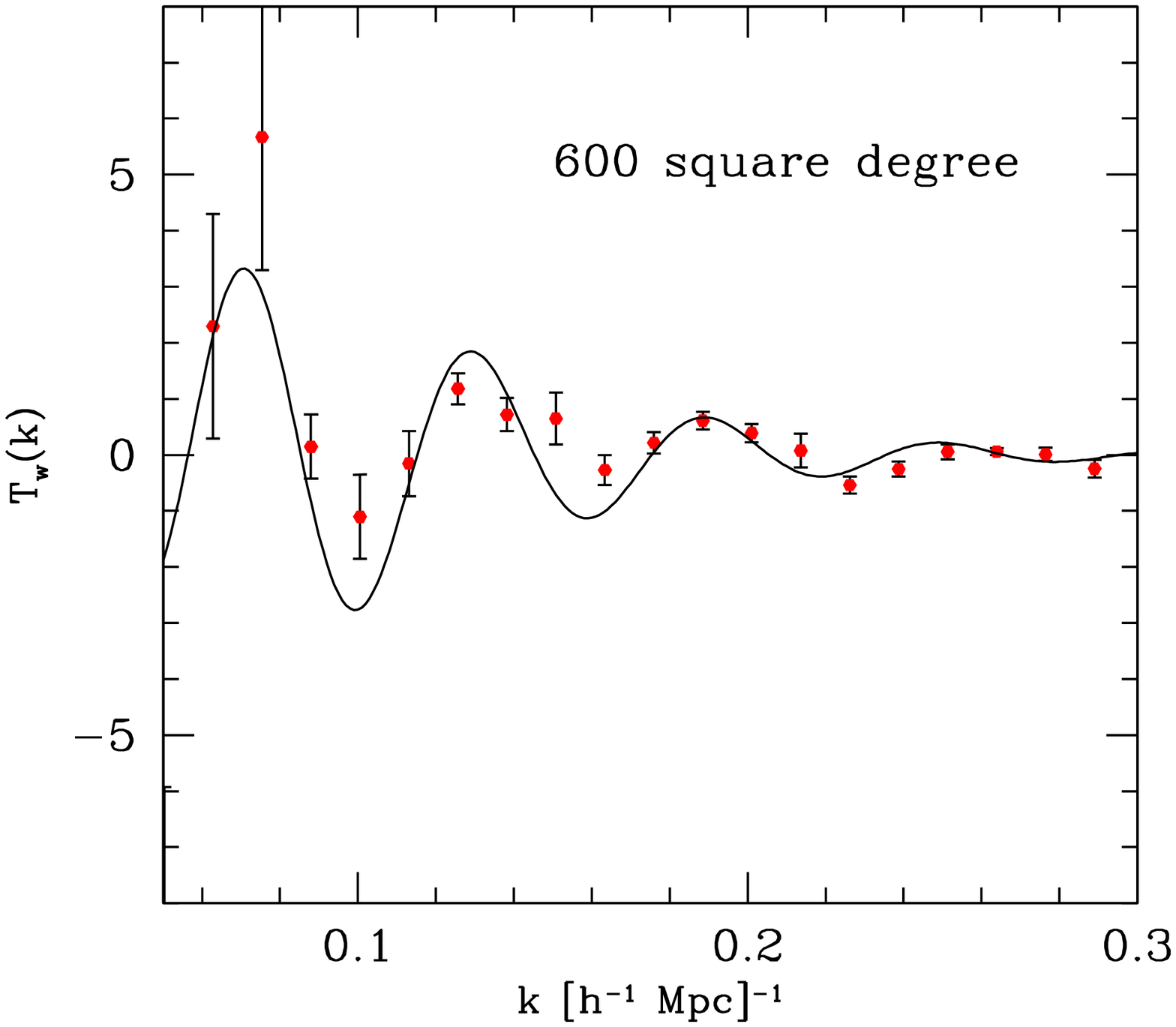}
    \caption[Combined wiggles]{\small Combined wiggle functions from
    the four Hubble Volume cubes. The upper left panel shows the
    combined data point (dots) with self-consistent error bars for
    the 100 square degree field. The upper right panel shows
    expected error bars for 200 square degrees and the lower row with
    error bars for 300 and 600 square degrees, respectively. The
    continuous lines are the theoretical wiggle functions for the
    reference (concordance) cosmology. } \label{fig combwiggles}
\end{figure}

Figure \ref{fig combwiggles} shows in the upper left panel the
combined wiggle function of the four XW wedge cubes of the Hubble
Volume Simulation. The combination is performed by averaging over
$k$-bins of discrete size. The errors represent the standard deviation
of all wiggle points in the corresponding $k$-bins. 
After averaging,
the 2nd to 5th BAOs are now clearly visible. We thus conclude that a
volume at least 4 times larger than SDSS has to be sampled in order to
see the BAOs with FITEX with high significance. One has to keep in mind, though,
that our cubes are located at much higher redshifts, where the smoothing effect of
non-linear structure growth is obviously much smaller. 
We also neglected the fact, that larger sky
coverages reduce the sizes of the fundamental modes, leading to 
higher (better) sampling rates of BAOs. 

With the 100 square degree sample, the possibility of data points
appearing at positions which severely inhibit the accuracy of the
cosmological test is relatively high, due to the high sample
variance. Especially as only a few data points, which are far away
from positions where theoretical wiggle functions of different
cosmological models intersect, have great statistical weight in the
cosmological test. This fact is shown in Fig.\,\ref{fig wigglescales}, where 
the modulus of the difference between the theoretical transfer functions 
for two models depending on position in $k$-space is plotted.

\begin{figure}[htbp]\hspace{-1.0cm}
    \centering \includegraphics[width=9cm]{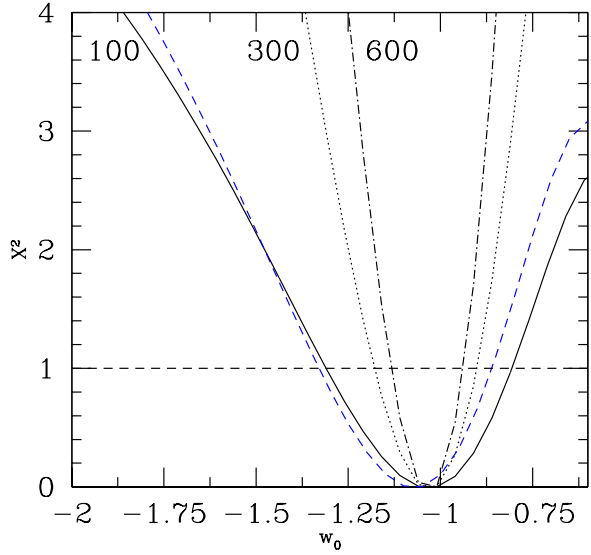}
    \vspace{-0.5cm}
    \caption[Cosmological test results]{\small $\chi^2$ distributions
    for the $w_0$ parameter obtained after marginalization over the
    amplitude parameter and for different conditions of BAO
    extraction. Short-dashed line: dark matter distribution in
    configuration space over 100 sq.\,deg obtained from the simulation
    of dark matter particles in configuration space. Continuous line:
    biased particle distribution in redshift space over 100
    sq.\,deg. Dotted line: biased particle distribution in redshift
    space over 300 sq.\,deg (extrapolation). Dashed-dotted line:
    biased particle distribution in redshift space over 600 sq.\,deg
    (extrapolation). The horizontal dashed line marks the $1\sigma$
    error level.}
\label{ERRORS}
\end{figure}

Figure \ref{ERRORS} shows the results of the cosmological test
obtained with BAOs extracted with FITEX under different survey
conditions. For surveys covering 100 square degrees we would expect errors of $\sigma_{w_0} = 0.25$.
The asymmetry of the error distributions directly reflects
the increased sensitivity of metric scales on $w$ for $w_0>-1$ compared
to $w_0<-1$. At these $w_0$ values, the theoretical wiggle function gets
quite compressed, so that at large $k$ values the amplitude decreases
rapidly due to Silk damping. When $w_0\rightarrow-\infty$, the
theoretical wiggle function formally mimics the case of zero
amplitude, so that low values of $w_0$ are excluded with about the same
significance as the ``no-wiggle'' case of zero amplitude. The minima
of all error distributions plotted in Fig.\,\ref{ERRORS} are slightly
off-set from the input $w_0=-1$ value, but well within all formal
$1\sigma$ ranges.

Note that the constraints given above are based solely on the phases of the
oscillations in the one-dimensional power spectrum, assuming perfect
knowledge of the sound horizon at drag epoch. 
The $w$-constraints are expected to further improve when the full two dimensional
redshift space effects as well as the amplitudes of the BAOs are used.

Perhaps the most important result of the present investigation is the
comparatively small decrease of only 8\% of $\sigma_{w_0}$ when comparing the
$w$ test with unbiased matter distributions in configuration
space, (dashed line in Fig. \ref{ERRORS}) i.e., the simplest survey 
condition, with the test that includes all aforementioned effects (solid line).
The robustness of the
FITEX method to extract BAOs from boosted transfer functions with
phenomenological fitting functions is mainly related to the fact that
only the amplitudes of the wiggle functions are boosted by
redshift-space and biasing effects, but the phases are not affected.
\\

To give an idea of the performance of FITEX based on surveys of larger size,
we extrapolated the results from the 100 square degree data.
The upper right panel of Fig.\,\ref{fig combwiggles} shows wiggle functions
extracted from the 100 square degree
data but with error bars expected for 200 - 600 square degrees. 
The error bars were extrapolated by scaling them according to the
expected number of additional $k$-modes. The different survey fields
are thus assumed to be not adjacent.

Figure\;\ref{fig combwiggles} shows that increasing the survey volume to 200 square degrees
does not reduce the error bars significantly, still allowing for crucial
data points at peaks of the oscillating wiggle function to be reduced
to zero amplitude. Error bars are still of the same order of magnitude
as the oscillations. Only at 300 square degrees (lower left panel),
the probability for smoothing out the oscillations due to statistical
errors gets considerably low. Following our extrapolations, for a
survey of 600 square degrees, the errors are expected to get much
smaller than the amplitude of the oscillations and the chance of
statistical deviations from the predicted function is of an order of
magnitude that allows for very accurate cosmological probing.

The accuracies of the $w_0$ parameter (Fig.\;\ref{ERRORS}) range
from $\sigma_{w_0}=0.25$ for 100 sq.\,deg which corresponds to 3 SDSS volumes, 
to $\sigma_{w_0}=0.09$ for 600
sq.\,deg, corresponding to 21 SDSS volumes. These results are obtained
for biased samples in redshift space, i.e., the most complex survey
condition. The improvement is directly related to the larger number of
independent $k$-modes in larger survey volumes and is consistent with results
obtained with Monte-Carlo simulations (Blake \& Glazebrook 2003, their Fig. 8).

Future studies should also investigate the various parameters of the cosmological test to get
constraints under less strong priors.

\section{Discussion and Conclusions}\label{DISCUSS}

We have seen that it is critical for the extraction of BAOs to either
fit a phenomenological function to the continuum of the boosted
transfer function (FITEX), or to calculate the shape of the boosted
transfer function, including all distortions down to the sub-percent
level. The latter approach has been carried out recently. Glazebrook
\& Blake (2005), H\"utsi (2005), and Springel et al. (2005) used 
a linear theory power spectrum and divided the `observed' data by this
reference. Almost the same approach, but with the addition of
linearizing the observational data before the division, to correct for
non-linear structure growth was used in Angulo et al. (2005). There
are two drawbacks of this method:

First, the ratio of a complex power spectrum and a reference power
spectrum does not resemble a wiggle function as defined in the text,
but something more complicated (see Eq.\,\ref{eq_division}). 
The simple form of the theoretical wiggle function (Eq.\,\ref{eq
wiggles}) is strongly deformed by both the chosen transfer function of
the reference spectrum and the distortions introduced by redshift
space and galaxy biasing. These distortions make the comparison with
theoretically expected model spectra unnecessarily complicated. FITEX
transforms the observed power spectrum into a boosted transfer
function where the subtraction of a non-oscillating phenomenological
continuum function is directly related to $T_w(k)$, multiplied by an
amplitude factor.

The second drawback is the linear theory power spectrum itself. By
computing a correct shape of the power spectrum, one has to rely
strongly on assumptions about the nature of the dark matter and
various other physical processes. The method is no more the `assumption
free' approach that was intended. Furthermore, the calculation of the
amplitude factor that includes all the distortions ranging from linear
structure growth, non-linear structure growth to redshift space
distortions and biasing is a problem. This calculation has to be
accurate to the sub-percent level, as the oscillations themselves make
up only $\sim\,2\%$ of the transfer function.

A more phenomenological approach was conducted by White (2005). He
assumed that redshift space distortions as well as linear and
non-linear structure growth could be calculated with reasonable
accuracy. Leaving only the ratio between dark matter power spectra and
`galaxy' power spectra, he found that one can fit this ratio with
about 2\% accuracy using a cubic polynomial. This adds further
evidence that a phenomenological approach can model power spectrum
distortions without assuming much about the physics of the distortions
themselves.

With the FITEX method, we have shown that one is able to extract the
BAOs without any strong priors on the nature of dark matter or the shape of
the power spectrum from the complex multi-component transfer
function. All the aforementioned effects are modeled by the continuum
function and need not be known very accurately, as long as the phases
of the oscillations are not distorted or washed out. This extraction
is accurate on the sub-percent level within $k$-ranges relevant for
the cosmological test. Furthermore, FITDEX is able to disentangle phase
information from amplitude information which could in the future be used to test the geometry
of the Universe on the one hand and the growth history on the other hand. This type of cosmological test could be used to discriminate Dark Energy from Modified Gravity theories.

\begin{acknowledgements}
We thank Ralf Bender for initiating this work on baryonic acoustic
oscillations, the HETDEX team, the Virgo Consortium for the provision
of the data of the Hubble Volume Simulation, Urs Seljak and Mathias
Zaldarriga for the CMBfast software, and Daniel Eisenstein and Wayne
Hu for the software to compute transfer functions for high baryon
fractions.
\end{acknowledgements}

\end{document}